# A repeated k-fold cross-validation approach for evaluating the instability of clinical prediction models: an empirical comparison to the bootstrap approach

Nop Khongthon[1], Pakpoom Wongyikul[1], Noraworn Jirattikanwong[1], Phanu Prasankittirach[1], Natthanaphop Isaradech[2], Wachiranun Sirikul[2], Noppadon Seesuwan[3], Suppachai Lawanaskol[4], Phichayut Phinyo[1]

[1] Department of Biomedical Informatics and Clinical Epidemiology (BioCE), Faculty of Medicine, Chiang Mai University, Chiang Mai, Thailand.

[2] Department of Community Medicine, Faculty of Medicine, Chiang Mai University, Chiang Mai, Thailand.

[3] Department of Emergency Medicine, Lampang Hospital, Lampang, Thailand.

[4] Chaiprakarn Hospital, Chiang Mai, Thailand.

**Correspondence:** Phichayut Phinyo; Department of Biomedical Informatics and Clinical Epidemiology (BioCE), Faculty of Medicine, Chiang Mai University, Chiang Mai, Thailand. Email: phichayut.phinyo@cmu.ac.th.

**Abstract**

**Background:** Bootstrap-based approaches have been recommended for assessing clinical prediction instability, but their performance relative to cross-validation (CV) remains unclear. We aimed to propose a CV-based approach for assessing prediction instability and compare it with a bootstrap-based approach across statistical and machine learning models.

**Methods:** We conducted a resampling-based empirical experiment using a real clinical cohort of 19,418 emergency department patients. Development samples were generated under events per variable (EPV) scenarios of 10, 30, and 50, and compared with the full dataset. Logistic regression (LR) and random forest (RF) models were evaluated using bootstrap validation and repeated 5-fold CV, with nested CV used for random forest hyperparameter tuning. Model performance was assessed using AUC, calibration slope, and calibration-in-the-large, while prediction instability was quantified using mean absolute prediction error (MAPE). A reference dataset was used to estimate empirical performance.

**Results:** For LR, bootstrap validation and 5-fold CV produced broadly similar discrimination and calibration, particularly at higher EPV values. For RF, apparent performance consistently overestimated the empirical AUC, defined as the AUC on the reference dataset. Bootstrap and 5-fold CV produced comparable discrimination. The 5-fold CV produced a calibration slope closer to the empirical value than bootstrap validation did. Prediction stability increased as EPV increased across both modelling approaches. At EPV 30, bootstrap-derived MAPE was higher than CV-derived MAPE for both LR and RF, with median MAPE values of 0.042 versus 0.020 and 0.077 versus 0.027, respectively.

**Conclusions:** A CV-based approach can assess prediction instability while also providing internal validated performance. Compared with bootstrap validation, 5-fold CV produced similar predictive performance for LR and calibration slope estimates closer to the empirical value for RF. CV-derived MAPE was consistently lower than bootstrap-derived MAPE, particularly for RF. These findings support the use of CV-based approaches for instability assessment, especially when comparing instability across multiple algorithms.

## 1. Background

Clinical prediction models (CPMs) are tools that employ statistical regression or machine learning methods to combine information from multiple predictors in order to estimate a clinical outcome of interest, either as an absolute risk prediction or as the expected value of a continuous clinical measurement (1-3). These predictions are used to assist or guide clinical decision-making in both diagnostic and prognostic contexts (4, 5). Each year, more than a thousand CPMs are developed in the biomedical field; however, only a small proportion are successfully validated and implemented in routine clinical practice (6, 7). While it is true that some models lack clinical utility (8), even those that are clinically useful still frequently exhibit methodological shortcomings that lead to overfitting. Consequently, their performance often fails to generalise to external or unseen data sets (7).

Several systematic reviews have found that many CPM studies continue to fall short of established methodological standards and exhibit practices that may be considered suboptimal, including the use of small development datasets, inappropriate predictor selection approaches, poor modelling choices, and improper handling of missing data (9-12). Such practices contribute not only to overfitting and instability in model performance metrics (reflecting the *validity* of the derived model), but also to instability in individual-level predictions (reflecting the *reliability* of model predictions) (11). To ensure that derived CPMs are sufficiently reliable and valid for clinical application since the development phase, internal validation and assessment of model instability should be undertaken (13, 14). This enables refinement of the modelling pipeline to reduce the risk of overfitting and instability before external evaluation.

Internal validation was originally proposed to assess the reproducibility of predictive performance by applying resampling approaches to the development dataset, such as split-sample, k-fold cross-validation, or bootstrapping methods (15). These approaches aim to quantify the degree of overfitting and to provide unbiased estimates of predictive performance under different resampling designs. Simulation studies have suggested that certain methods may perform better than others depending on the data structure, dimensionality, and complexity of the modelling context (15-19). The choice of internal validation approach also varies considerably across the literature and is often influenced by disciplinary preferences (19, 20), with cross-validation more commonly used in machine learning applications and bootstrap more frequently adopted in statistical modelling (12, 21).

Recently, the bootstrap approach has been recommended as a standard approach for evaluating the instability of CPM predictions (13), as it closely resembles the generation of multiple samples from the same target population to develop prediction models using an identical modelling pipeline and subsequently generate predictions for the same individuals (22). This approach allows straightforward calculation of instability metrics, such as the mean absolute prediction error (MAPE), and facilitates assessment of variation in predictions across resampled datasets. However, whether the bootstrap approach provides a valid and unbiased assessment of prediction instability, compared with alternative approaches such as cross-validation, has not been formally examined. This gap is particularly relevant in the context of complex modelling approaches involving machine learning with hyperparameter tuning, where data leakage is one of the concerns (3, 23) and resampling approaches must appropriately replicate the full modelling pipeline to avoid biased instability estimates.

In this study, we aim to propose alternative cross-validation-based approach for evaluating prediction instability and to conduct a comparative evaluation against the reference bootstrap approach in both statistical modelling and machine learning settings.

## 2. Methods

### 2.1. Study design

This was a resampling-based empirical experiment using a real clinical cohort to evaluate prediction instability. We proposed a cross-validation-based approach for assessing prediction instability and compared it with the bootstrap-based approach proposed by Riley et al. Both approaches were applied to modelling algorithms with and without hyperparameter tuning and evaluated across varying sample size conditions defined by events-per-variable (EPV). Although prediction instability was the primary focus of this study, model performance was assessed first to provide context for interpreting the instability results. This ordering reflects the conventional sequence of prediction model evaluation, in which discrimination and calibration are examined before assessing whether predictions remain stable across resamples.

### 2.2. Data source

The data used in this study were derived from a previously published clinical prediction model (CPM) developed to predict hospital admission among patients presenting to the emergency department (ED) at Lampang Hospital, Thailand. The dataset comprised 19,418 ED patients, of whom 7,831 (40.3%) were admitted. Rather than using simulated data, we employed a repeated subsampling approach applied to the real cohort. This approach preserves the observed covariate distributions, event rate, and predictor–outcome associations within the source dataset, while allowing controlled variation in sample size, expressed as events-per-variable (EPV), to evaluate its impact on model performance and prediction instability across validation approaches.

### 2.3. Model development pipelines

To develop prediction model for hospitalisation, we pre-defined six candidate predictors: age (years), mean arterial pressure (MAP, mmHg), pulse rate (PR, bpm), respiratory rate (RR, breaths/min), peripheral oxygen saturation ($SpO_2$, %), and body temperature (BT, °C). All predictors were assumed to be linearly associated with the outcome. As the dataset contained complete data for all predictors, no missing data handling was required.

As we intended to assess whether the performance of different instability evaluation approaches would vary according to the type of model fitted (with or without hyperparameter tuning), we defined two modelling algorithms: (1) a binary logistic regression (LR) model, including all six predictors as main effects without penalisation or variable selection, and (2) a random forest (RF) model using the *ranger* engine within the *tidymodels* framework in R. Three hyperparameters were tuned: minimum node size (range: 20–30), number of trees (range: 200–300), and maximum tree depth (range: 3–6). A space-filling grid of 100 candidate hyperparameter combinations was generated prior to model fitting. The optimal hyperparameter combination within each resampling approach was selected by maximising the area under the receiver operating characteristic curve, evaluated on held-out data.

### 2.4. Internal validation and instability evaluation

After a model has been developed, it should be evaluated on unseen data to obtain and accurate estimate of its true predictive performance (θ). In the absence of external test data, resampling methods such as cross-validation (CV) and bootstrap resampling of the training data are generally recommended to provide an unbiased performance estimate. This process is called internal validation. Recently, bootstrap resampling has also been proposed for assessing prediction instability. Details for each approach in this study are as follows:

### ***2.4.1. Bootstrap-based approach***

This bootstrap evaluation followed the approach proposed by Riley et al. The development sample was resampled $B$ =200 times with replacement. For each bootstrap resample b=1,…, $B$, the model or algorithm was fitted on the bootstrap sample and evaluated on both the bootstrap sample itself, $\theta_{\text{app},b}$ (apparent performance), and the original development sample, $\theta_{\text{boot},b}$ (bootstrap performance). For algorithms requiring tuning (e.g., random forest), hyperparameters were selected within each bootstrap resample using internal 5-fold CV, after which the model was refitted on the same bootstrap sample. The apparent ($\theta_{\text{app},b}$) and bootstrap performance ($\theta_{\text{boot},b}$) estimates were used to derive the optimism-corrected performance ($\theta_{\text{corrected}}$).

$$\text{optimism} = \frac{1}{B}\sum_{b=1}^{B}\left(\theta_{\text{app},b} - \theta_{\text{boot},b}\right)$$

$$\theta_{\text{corrected}} = \theta_{\text{origi}} - \text{optimism}$$

The model fitted within each bootstrap resample, referred to as the bootstrap model, was also used to generate predicted probabilities ($\widehat{p_i^{(b)}}$) for each individual $i$ in both the bootstrap sample and the original development sample. These predictions were subsequently used to assess prediction instability, as described later.

### *2.4.2. Cross-validation-based approach*

Cross-validation was implemented according to whether the model required hyperparameter tuning. Models without tuning were evaluated using repeated CV, whereas models requiring tuning were evaluated using repeated nested CV. We adopted a 5-fold CV with 200 repeats. In repeated $K$ -fold CV, the development sample is randomly partitioned into $K$ approximately equal parts, with one part held out for model evaluation and the remaining $K-1$ parts used for model development. This process is repeated $K$ times so that each part serves once as the validation set, yielding $K$ performance estimates per repeat. These estimates, denoted $\theta_{k,r}$, are averaged across all folds and repeats to obtain the cross-validated performance:

$$\theta_{\mathrm{CV}} = \frac{1}{KR}\sum_{r=1}^{R}\sum_{k=1}^{K}\theta_{k,r}$$

where $K$ is the number of folds and $R$ is the number of repeats. Under standard assumptions, the average cross-validated performance provides an approximately unbiased estimate of test performance and is theoretically comparable to optimism-corrected performance ($\theta_{\mathrm{corrected}}$).

For each repeat $r=1,\ldots,R$, $K$ models were developed using $K-1$ partitions and evaluated on the remaining partition. Each model generated predicted probabilities ($\widehat{p_i^{(k,r)}}$) for individuals $i$ in its corresponding held-out partition $k=1,\ldots,K$. Consequently, across the $K$ folds, all individuals in the dataset received one out-of-sample prediction per repeat. These predictions were subsequently used to assess prediction instability.

For models requiring hyperparameter tuning, repeated nested $K$-fold CV was implemented. In this approach, the outer loop followed the same $K$-fold structure described above to provide an unbiased estimate of model performance. Within each outer training partition, an inner $K$-fold

CV loop was used to select optimal hyperparameters. Specifically, for each outer fold and repeat, candidate hyperparameter configurations were evaluated using CV within the training data, and the optimal configuration, based on the mean performance across inner folds, was then used to fit the model, which was subsequently evaluated on the held-out outer partition. This process was repeated across all folds and repeats.

### 2.5. Simulation and scenario specifications

The simulation started with a stratified random sample, preserving the outcome proportion, corresponding to a specified events-per-variable (EPV) size of 30 drawn from the full dataset (n = 19,418). EPV sizes of 10 and 50, together with the full dataset at an EPV of 1004, were also included to assess trends across EPV levels. This sample was then randomly split into a simulated dataset (80%) and a reference dataset (20%). The simulated dataset was used for model development and internal validation, whereas the reference dataset was reserved as a pseudo-external validation dataset and was not involved in any model development, tuning, or selection.

Under standard bootstrap internal validation and instability evaluation, the full dataset would typically be used for model development. However, as the present study aimed to compare bootstrap with 5-fold CV, where in each iteration models are developed using 80% (4/5 folds) of the data, the effective development sample size was aligned across approaches. Accordingly, for bootstrap-based analyses, 20% of the simulated dataset was set aside to mimic common real-world practice in which data are split into training and test sets, and models were developed using the remaining 80%, with bootstrap resampling performed within this subset. In contrast, for 5-fold CV, the full simulated dataset was used, with the 80/20 split arising within each fold.

Predictive performance in the reference dataset, representing an unseen pseudo-external dataset, was regarded as the model's empirical performance, defined as the benchmark obtained on the reference dataset against which the internally validated estimates were compared. In each simulation, a single prediction model was developed and evaluated on the reference dataset. A total of 200 independent simulation iterations were performed. In each iteration, a development subsample of a specified size was randomly drawn without replacement from the full dataset, and a fixed random seed, with iteration- and condition-specific offsets, was used to ensure reproducibility across all runs. Figure 1 and Figure 2 illustrates the analytical flow for evaluating the predictive performance and the prediction instability, respectively.

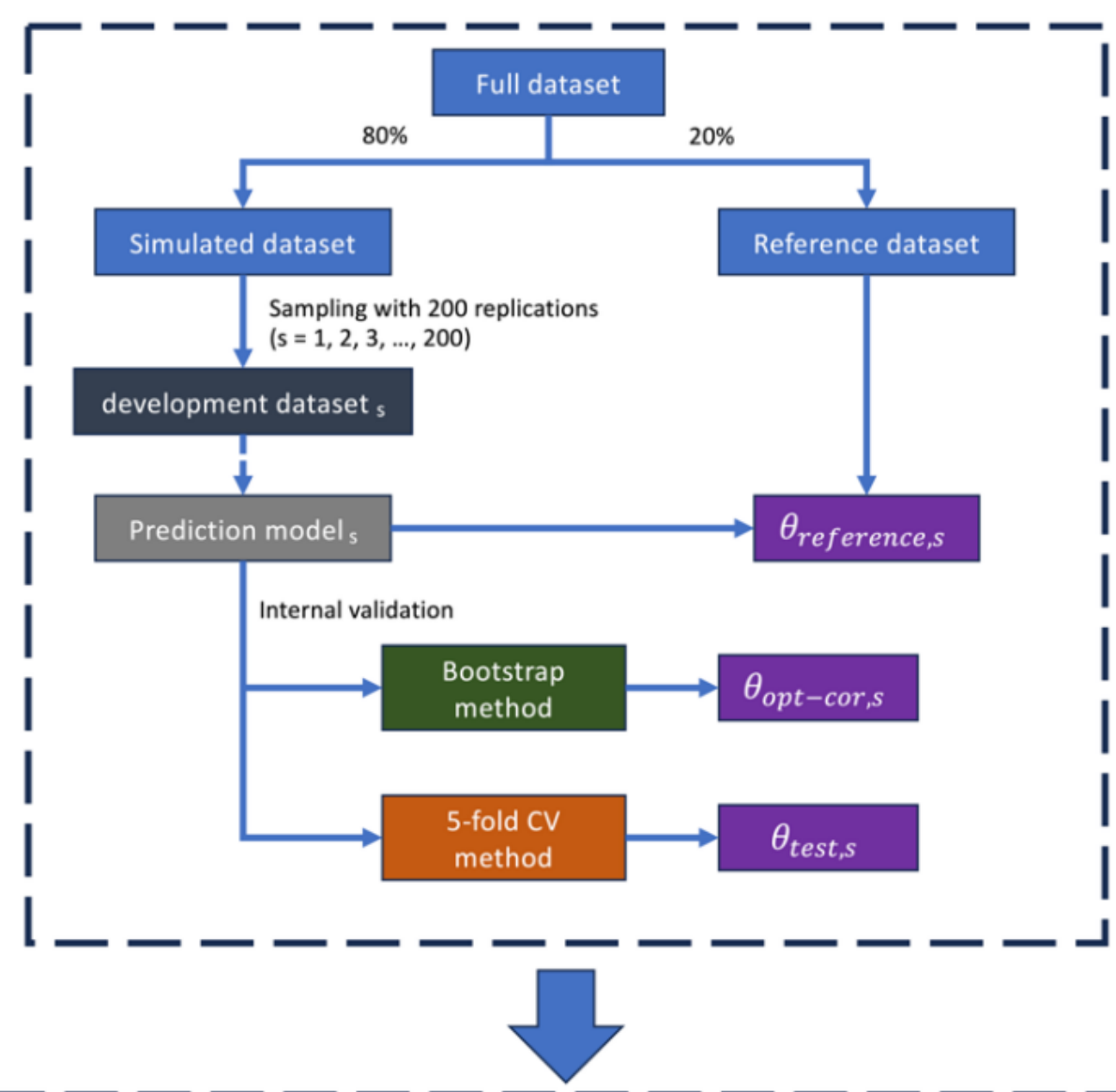
Full dataset
80%
20%
Simulated dataset
Reference dataset
Sampling with 200 replications
(s = 1, 2, 3, …, 200)
development dataset s
Prediction model s
$\theta_{reference,s}$
Internal validation
Bootstrap method
$\theta_{opt-cor,s}$
5-fold CV method
$\theta_{test,s}$

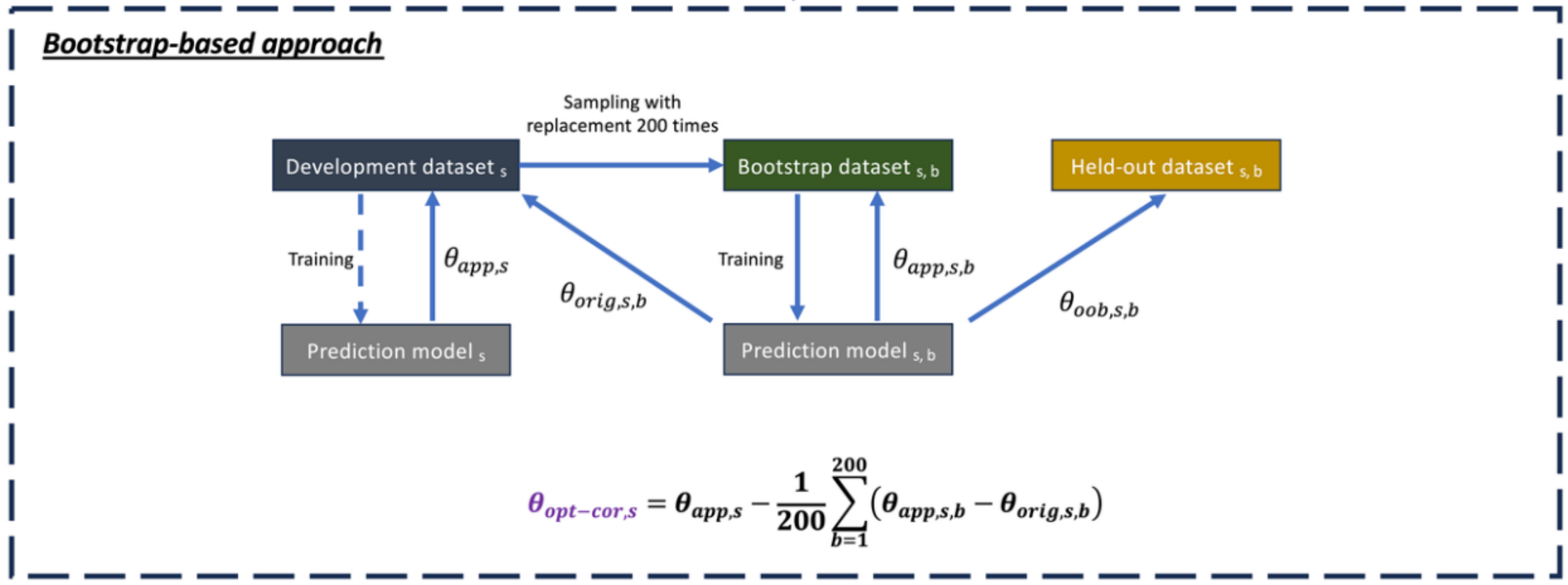
Bootstrap-based approach
Sampling with replacement 200 times
Development dataset s
Bootstrap dataset s, b
Held-out dataset s, b
Training
$\theta_{app,s}$
$\theta_{orig,s,b}$
Training
$\theta_{app,s,b}$
$\theta_{oob,s,b}$
Prediction model s
Prediction model s, b
$\theta_{opt-cor,s} = \theta_{app,s} - \frac{1}{200}\sum_{b=1}^{200}(\theta_{app,s,b} - \theta_{orig,s,b})$

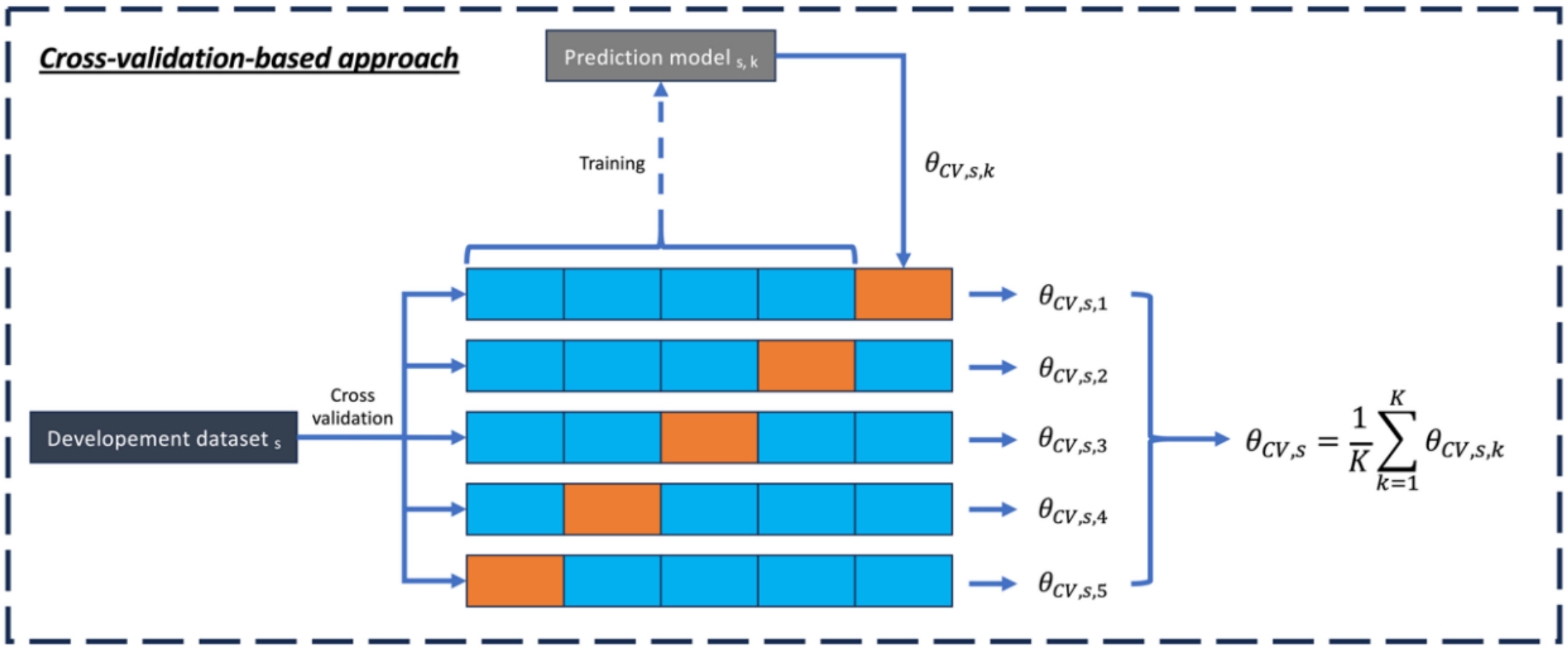
Cross-validation-based approach
Prediction model s, k
Training
$\theta_{CV,s,k}$
Developement dataset s
Cross validation
$\theta_{CV,s,1}$
$\theta_{CV,s,2}$
$\theta_{CV,s,3}$
$\theta_{CV,s,4}$
$\theta_{CV,s,5}$
$\theta_{CV,s} = \frac{1}{K}\sum_{k=1}^{K}\theta_{CV,s,k}$

**Figure 1. Analytical flow for evaluating the predictive performance across simulated datasets.** The full dataset was split into a simulated dataset and a reference dataset. Development datasets were generated by resampling the simulated dataset 200 times (s = 1, 2, …, 200), each used to fit a prediction model and evaluated on the reference dataset to obtain reference performance. Internal validation was performed using two approaches. In the bootstrap-based approach, each development dataset was resampled with replacement 200 times; apparent performance and performance on the original development dataset were obtained per iterations, and optimism-corrected performance was computed as the mean difference between the two. In the cross-validation-based approach, each development dataset was partitioned into 5 folds; each fold served once as the validation set, and mean performance was averaged across all folds. **Abbreviations**: app, apparent; CV, cross-validation; OOB, out-of-bag; opt-cor, optimism-corrected; s, simulation index; b, bootstrap iterations index; k, fold index.

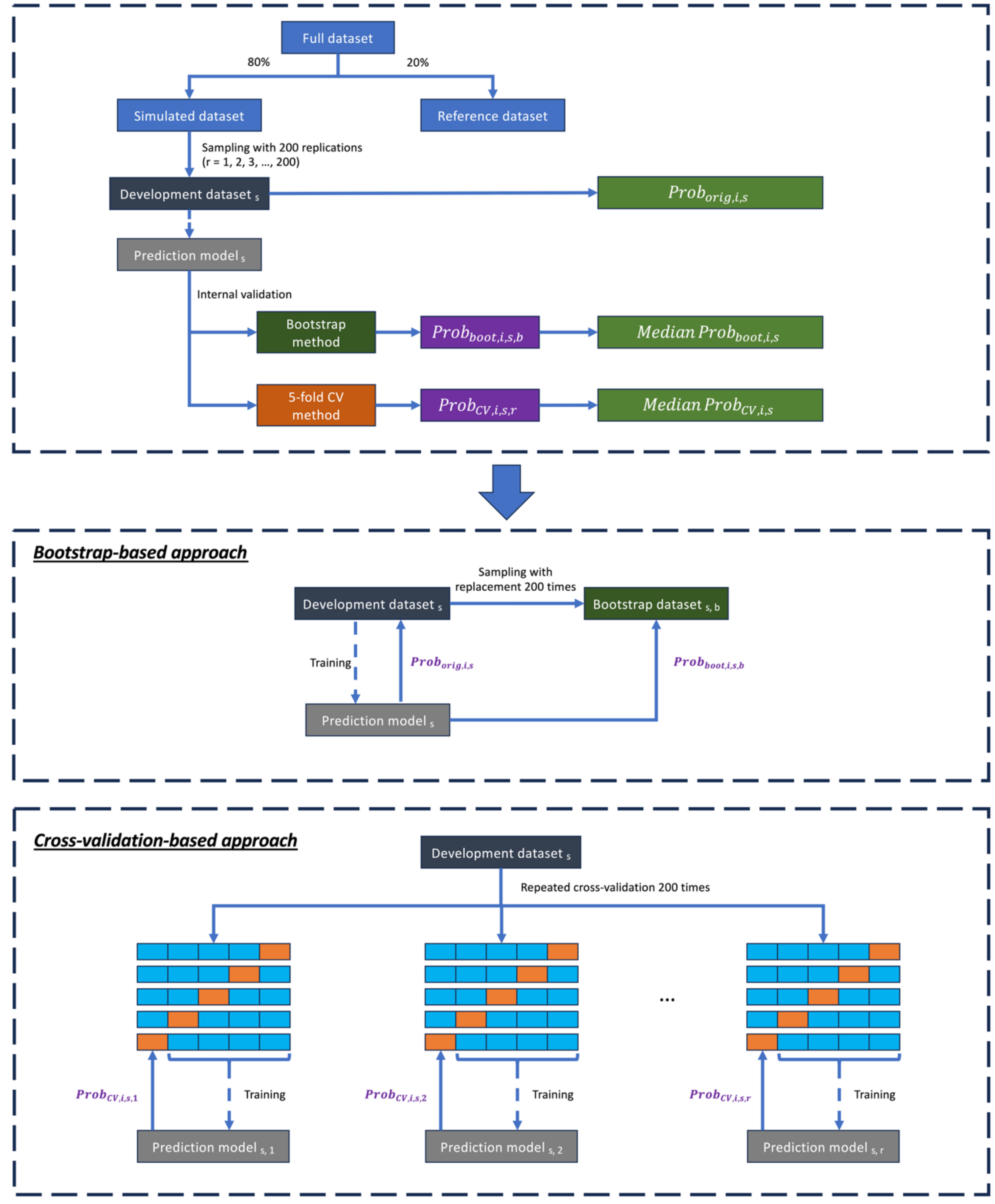

**Figure 2. Analytical flow for evaluating the prediction instability across simulated datasets.** The full dataset was split into a simulated dataset and a fixed reference dataset. Development datasets were generated by resampling the simulated dataset 200 times (s = 1, 2, …, 200), each used to fit a prediction model. Original predicted probabilities from each model served as the reference for instability assessment. In the bootstrap-based approach, each development dataset was resampled with replacement 200 times; the fitted model per iterations was applied back to the original development dataset, and the median predicted probability across 200 iterations was derived per individual. In the cross-validation-based approach, 5-fold cross-validation was repeated 200 times; fold-specific models generated predicted probabilities for held-out individuals, and the median across 200 repetitions was derived per individual. Prediction instability and MAPE were computed from the deviation of iteration-specific predicted probabilities around their respective reference probabilities across all simulated samples. **Abbreviations**: CV, cross-validation; MAPE, mean absolute prediction error; orig, original; b, bootstrap iterations index; s, simulation index; r, repeated cross-validation index.

### 2.6. Instability measures

The instability of model predictions was quantified using the mean absolute prediction error (MAPE). This measure was calculated differently for bootstrap and cross-validation (CV) approaches. For bootstrap-based MAPE, the calculation followed the approach described by Riley et al. For each individual $i$, the individual-level MAPE was defined as the mean absolute difference between the predicted probability from each bootstrap model and the predicted probability from the model fitted on the original development sample:

$$\mathrm{MAPE_i} = \frac{1}{B}\sum_{b=1}^{B}\left|\widehat{p_{i,b}^{\mathrm{boot}}} - \widehat{p_i^{\mathrm{orig}}}\right|$$

where $\hat{p}_{i,b}^{\mathrm{boot}}$ denotes the predicted probability for individual $i$ from bootstrap resample $b$, $\hat{p}_i^{\mathrm{orig}}$ denotes the predicted probability from the model fitted on the original development sample, and $B$ is the total number of bootstrap resamples. The average MAPE across all individuals was then calculated as:

$$\text{MAPE} = \frac{1}{BN}\sum_{i=1}^{N}\sum_{b=1}^{B}\left|\widehat{p_{i,b}^{\text{boot}}} - \widehat{p_i^{\text{orig}}}\right|$$

where $N$ is the total number of individuals.

For cross-validation (CV)-based MAPE, prediction instability was assessed across repeated K-fold cross-validation. In each repeat, the dataset was randomly partitioned into $K$ folds, with models trained on $K-1$ folds and evaluated on the remaining fold. Consequently, within each repeat, each individual received a single out-of-sample predicted probability from the model corresponding to the fold in which they were held out. Across $R$ repeats, this resulted in $R$ predicted probabilities for each individual, each obtained from a different model trained on a different subset of the data. For each individual $i$, the individual-level MAPE was defined as the mean absolute deviation of these predicted probabilities from the median predicted probability across all repeats:

$$\text{MAPE}_{\text{i}}^{\text{CV}} = \frac{1}{R}\sum_{r=1}^{R}\left|\widehat{p_{i,r}} - \widetilde{p_i}\right|$$

where $\hat{p}_{i,r}$ denotes the predicted probability for individual $i$ in repeat $r$, and $\tilde{p}_i$ denotes the median predicted probability for individual $i$ across all $R$ repeats. The average MAPE across all individuals was calculated as:

$$\text{MAPE}^{\text{CV}} = \frac{1}{N}\sum_{i=1}^{N} \text{MAPE}_{\text{i}}^{\text{CV}}$$

This definition differs from bootstrap-based MAPE. In the bootstrap setting, predictions from each resampled model can be directly compared with those from a single reference model fitted on the original development sample. In contrast, under repeated cross-validation, predictions are generated from multiple models trained on different subsets of the data, and no single reference model exists that provides comparable predictions for all individuals. Therefore, the median prediction across repeats was used as the reference.

Other aspects of prediction instability were also examined using visualisations, following the approach proposed by Riley et al., including MAPE instability plots, prediction instability plots, and calibration instability plots. As a supplementary analysis, bootstrap-based MAPE and the corresponding instability plots were additionally computed against the median predicted probability across resamples, in the same form as the cross-validation definition, so that the two approaches could be compared on a common footing (Supplementary Figures S5 and S7).

**2.7. Performance measures**

The model performance was assessed in terms of discrimination and calibration. Discrimination was evaluated using the area under the receiver operating characteristic curve (AUC), estimated non-parametrically, with values closer to 1.0 indicating better ability to distinguish admitted from non-admitted patients. Calibration was assessed using two metrics. The calibration slope (ideal value = 1.0) quantifies the agreement between predicted and observed risks across the range of predictions; values below 1.0 indicate overfitting, whereas values above 1.0 indicate underfitting. Calibration-in-the-large (ideal value = 0) quantifies systematic over- or underprediction at the mean level. Calibration curves were generated for a representative iteration, defined as the iteration whose mean absolute prediction error was closest to the median across all 200 iterations within each approach and EPV condition.

### 2.8. Statistical analysis and software

Results from all iterations were summarised descriptively using medians with 95% empirical percentile intervals (2.5th–97.5th). These intervals reflect variability across simulation iterations rather than inferential uncertainty. Comparisons focused on differences in prediction performance and instability metrics without formal hypothesis testing. All analyses were conducted in R (version 4.5.2) using the *tidymodels* framework and the *ranger*, *ggplot2*, *gtsummary*, and *gt* packages. All scripts were executed as array jobs on a high-performance computing (HPC) cluster at Chiang Mai University. Simulation code and analysis scripts are available at: https://github.com/Nopkhongthon/predinstab.

## 3. Results

### Study datasets

The full dataset comprised 19,418 observations, with an outcome incidence of 40.3%. The predictor and outcome distributions were highly similar between the simulated and reference datasets. Table 1 presents the detailed characteristics of the full dataset and the distribution of predictors in the simulated and reference datasets.

**Table 1:** Baseline characteristics of the full dataset and a sample of the simulated and the reference dataset.

| Characteristics | Full dataset (n = 19418) | Simulated dataset (n = 15533)* | Reference dataset (n = 3885)* | STD |
|---|---|---|---|---|
| Age, years | 49.6 (22.0) | 49.44 (22.02) | 50.17 (21.71) | -0.034 |
| Mean arterial pressure, mmHg | 98.0 (17.3) | 98.04 (17.30) | 97.86 (17.07) | 0.011 |
| Pulse rate, bpm | 90.1 (19.1) | 90.14 (19.08) | 89.82 (19.01) | 0.017 |
| Respiratory rate, /min | 20.3 (3.4) | 20.28 (3.42) | 20.32 (3.40) | -0.011 |
| $SpO_2$, % | 97.5 (3.3) | 97.46 (3.36) | 97.48 (3.09) | -0.006 |
| Temperature, °C | 36.8 (0.8) | 36.73 (0.83) | 36.72 (0.82) | 0.017 |
| Admission (outcome), n (%) | 7831 (40.3%) | 6264 (40.3%) | 1567 (40.3%) | -0.000 |

[1] Mean (± SD), n (%). * Values presented are representative of simulated dataset 1.

**Abbreviations**: bpm, beat per minute; SD, standard deviation; $SpO_2$, oxygen saturation; STD, standardization difference.

## Predictive performance

For logistic regression, AUC estimates were similar across validation approaches at higher EPV values and converged towards the empirical AUC. At low EPV, the apparent AUC overestimated empirical performance, while bootstrap and 5-fold CV slightly underestimated it. Calibration slope and CITL were generally close to the empirical values, particularly at higher EPV. At lower EPV, bootstrap estimates showed greater variability, whereas CV appeared more stable. For random forest, the apparent AUC consistently overestimated empirical performance, including in the full dataset. At higher EPV, bootstrap and 5-fold CV gave similar AUC estimates, although bootstrap estimates were more variable. For calibration slope, CV was closer to the empirical value than bootstrap at higher EPV, with both approaches becoming similar in the full dataset. CITL was close to 0 overall, but bootstrap correction showed greater variability, while CV was more stable and produced minimally higher estimates. Table 2 and Figure 3 present and compare the performance metrics across EPV scenarios for both validation approaches.

**Table 2: Prediction performance metrics across approaches.** Values are presented of internally validated predictive performances included AUC, calibration slope, CITL as median (95% empirical percentile intervals (2.5$^{th}$–97.5$^{th}$)) across simulation iterations. Empirical prediction performance evaluated on the reference sample.

| EPV | Logistic regression | | | Random forest | | |
|---|---|---|---|---|---|---|
| | Apparent | Bootstrap | 5-fold CV | Apparent | Bootstrap | 5-fold CV |
| **AUC** | | | | | | |
| EPV 10 | 0.738 (0.653, 0.815) | 0.697 (0.596, 0.789) | 0.689 (0.590, 0.779) | 0.869 (0.813, 0.935) | 0.752 (0.668, 0.836) | 0.682 (0.562, 0.761) |
| EPV 30 | 0.726 (0.669, 0.769) | 0.712 (0.650, 0.758) | 0.710 (0.647, 0.757) | 0.803 (0.764, 0.889) | 0.703 (0.643, 0.801) | 0.710 (0.639, 0.761) |
| EPV 50 | 0.721 (0.689, 0.755) | 0.713 (0.678, 0.748) | 0.711 (0.677, 0.746) | 0.790 (0.749, 0.862) | 0.703 (0.646, 0.783) | 0.716 (0.675, 0.754) |
| Full | 0.719 (0.716, 0.723) | 0.719 (0.716, 0.723) | 0.719 (0.716, 0.722) | 0.747 (0.744, 0.751) | 0.732 (0.729, 0.736) | 0.732 (0.729, 0.736) |
| Empirical | 0.719 (0.704, 0.731) | | | 0.727 (0.712, 0.740) | | |
| **Calibration slope** | | | | | | |

| EPV | Logistic regression | | | Random forest | | |
|---|---|---|---|---|---|---|
| | Apparent | Bootstrap | 5-fold CV | Apparent | Bootstrap | 5-fold CV |
| EPV 10 | 1.000 (1.000, 1.000) | 0.780 (0.623, 0.857) | 0.864 (0.611, 1.020) | 2.371 (1.813, 3.331) | 0.796 (0.339, 1.493) | 1.185 (0.652, 1.599) |
| EPV 30 | 1.000 (1.000, 1.000) | 0.917 (0.870, 0.942) | 0.949 (0.886, 0.977) | 1.906 (1.646, 2.424) | 0.707 (0.267, 1.014) | 1.351 (1.177, 1.520) |
| EPV 50 | 1.000 (1.000, 1.000) | 0.948 (0.928, 0.968) | 0.968 (0.945, 0.983) | 1.754 (1.567, 2.056) | 0.776 (0.517, 1.038) | 1.398 (1.242, 1.544) |
| Full | 1.000 (1.000, 1.000) | 0.998 (0.995, 1.001) | 0.998 (0.998, 0.999) | 1.234 (1.221, 1.247) | 1.122 (1.110, 1.135) | 1.220 (1.211, 1.231) |
| Empirical | 0.999 (0.898, 1.092) | | | 1.225 (1.110, 1.339) | | |
| **CITL** | | | | | | |
| EPV 10 | 0.000 (-0.000, 0.000) | 0.008 (-0.023, 0.039) | 0.005 (-0.014, 0.025) | 0.003 (-0.018, 0.023) | 0.008 (-0.029, 0.042) | 0.013 (-0.007, 0.036) |
| EPV 30 | 0.000 (-0.000, 0.000) | 0.001 (-0.014, 0.016) | 0.001 (-0.002, 0.006) | 0.001 (-0.009, 0.011) | 0.007 (-0.010, 0.027) | 0.012 (0.005, 0.020) |

| EPV | Logistic regression | | | Random forest | | |
|---|---|---|---|---|---|---|
| | Apparent | Bootstrap | 5-fold CV | Apparent | Bootstrap | 5-fold CV |
| EPV 50 | 0.000 (0.000, 0.000) | 0.000 (-0.010, 0.013) | 0.001 (-0.001, 0.003) | 0.001 (-0.005, 0.011) | 0.010 (-0.004, 0.021) | 0.010 (0.005, 0.016) |
| Full | 0.000 (0.000, 0.000) | 0.000 (-0.001, 0.001) | 0.000 (0.000, 0.000) | 0.002 (0.001, 0.002) | 0.003 (0.001, 0.004) | 0.003 (0.003, 0.003) |
| Empirical | -0.001 (-0.034, 0.028) | | | 0.000 (-0.027, 0.025) | | |

Abbreviations: AUC, area under the receiver operating characteristic curve; CITL, calibration-in-the-large; EPV, events-per-variable; Apparent, performance estimated on the same data used for model development; Bootstrap, internal validation using the bootstrap approach; 5-fold CV, internal validation using 5-fold cross-validation; Empirical, performance evaluated on the reference sample.

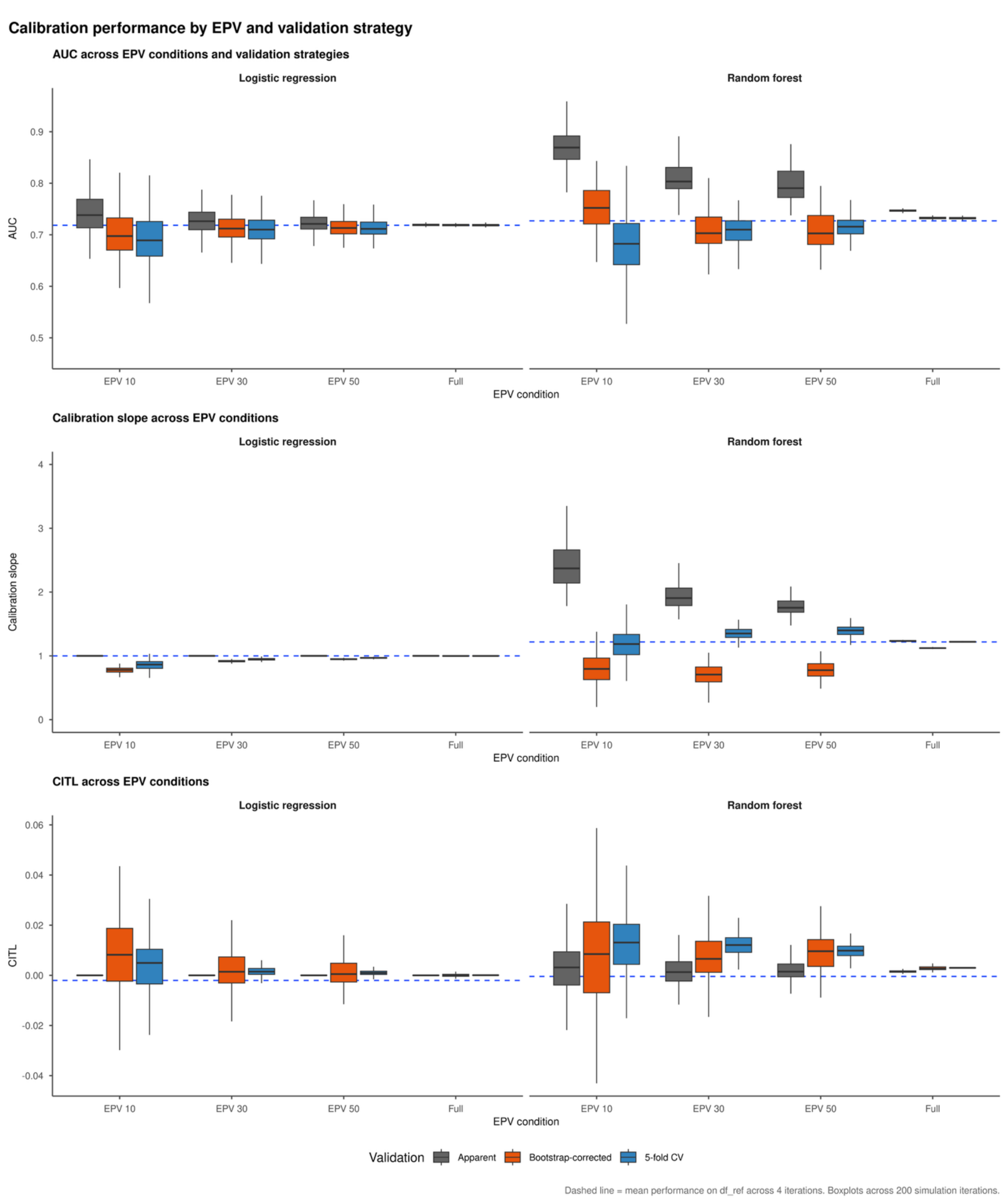

**Figure 3. Predictive performance by EPV condition and validation approach across logistic regression and random forest models.** Each row displays a different performance metric: area under the receiver operating characteristic curve (AUC, upper panels), calibration slope (middle

panels), and Calibration-in-the-large (CITL, lower panels). Left and right panels correspond to logistic regression and random forest models, respectively. The x-axis represents the events-per-variable (EPV) condition (EPV 10, EPV 30, EPV 50, and full dataset). Boxplots display the distribution of performance estimates across 200 simulated samples for each validation approach: apparent (grey), bootstrap (orange), and 5-fold cross-validation (blue). The horizontal dashed line indicates the mean performance on the reference dataset, serving as the benchmark against which validation estimates are compared. **Abbreviations**: AUC, area under the receiver operating characteristic curve; CITL, Calibration-in-the-large; CV, cross-validation; EPV, events-per-variable.

**Prediction instability**

Prediction stability increased as EPV increased, with the lowest MAPE observed in the full dataset for both modelling and validation approaches (Table 3). At EPV 30, bootstrap-derived MAPE was higher than 5-fold CV-derived MAPE for both logistic regression and random forest. For logistic regression, the median MAPE was 0.042 using bootstrap validation compared with 0.020 using 5-fold CV. For random forest, the corresponding values were 0.077 and 0.027, respectively. At the same EPV level, random forest showed higher MAPE than logistic regression, particularly for bootstrap-derived MAPE. In contrast, the difference between logistic regression and random forest was smaller for CV-derived MAPE. This matches the reference dataset results, where bootstrap was more optimistic than 5-fold CV for random forest. These patterns are illustrated in Figure 4. Corresponding instability plots for the remaining events-per-variable conditions, sample size scenarios, and modelling approaches are provided in Additional file 1.

**Table 3: Average mean absolute prediction error (MAPE) across modelling approaches.** Values are presented of MAPE as median (95% empirical percentile intervals (2.5th–97.5th)) across simulation iterations.

| **Dataset scenario** | **Average mean absolute prediction error (MAPE)** median (95% empirical percentile intervals (2.5th–97.5th) | | | |
|---|---|---|---|---|
| | **reg_boot** | **reg_cv** | **rf_boot** | **rf_cv** |
| EPV 10 | 0.075 | 0.036 | 0.098 | 0.038 |
| | (0.067, 0.084) | (0.033, 0.041) | (0.087, 0.106) | (0.033, 0.043) |
| EPV 30 | 0.042 | 0.020 | 0.077 | 0.027 |
| | (0.040, 0.045) | (0.019, 0.021) | (0.072, 0.082) | (0.022, 0.034) |
| EPV 50 | 0.033 | 0.015 | 0.066 | 0.025 |
| | (0.031, 0.034) | (0.014, 0.016) | (0.062, 0.070) | (0.018, 0.030) |
| Full | 0.007 | 0.003 | 0.023 | 0.010 |
| | (0.007, 0.007) | (0.003, 0.003) | (0.022, 0.023) | (0.009, 0.010) |

**Abbreviations**: EPV, events-per-variable; reg_boot, logistic regression with an internal validation using the bootstrap approach; reg_cv, logistic regression with an internal validation using the 5-fold cross-validation approach; rf_boot, random forest with an internal validation using the bootstrap approach; rf_cv, random forest with an internal validation using the 5-fold cross-validation approach.

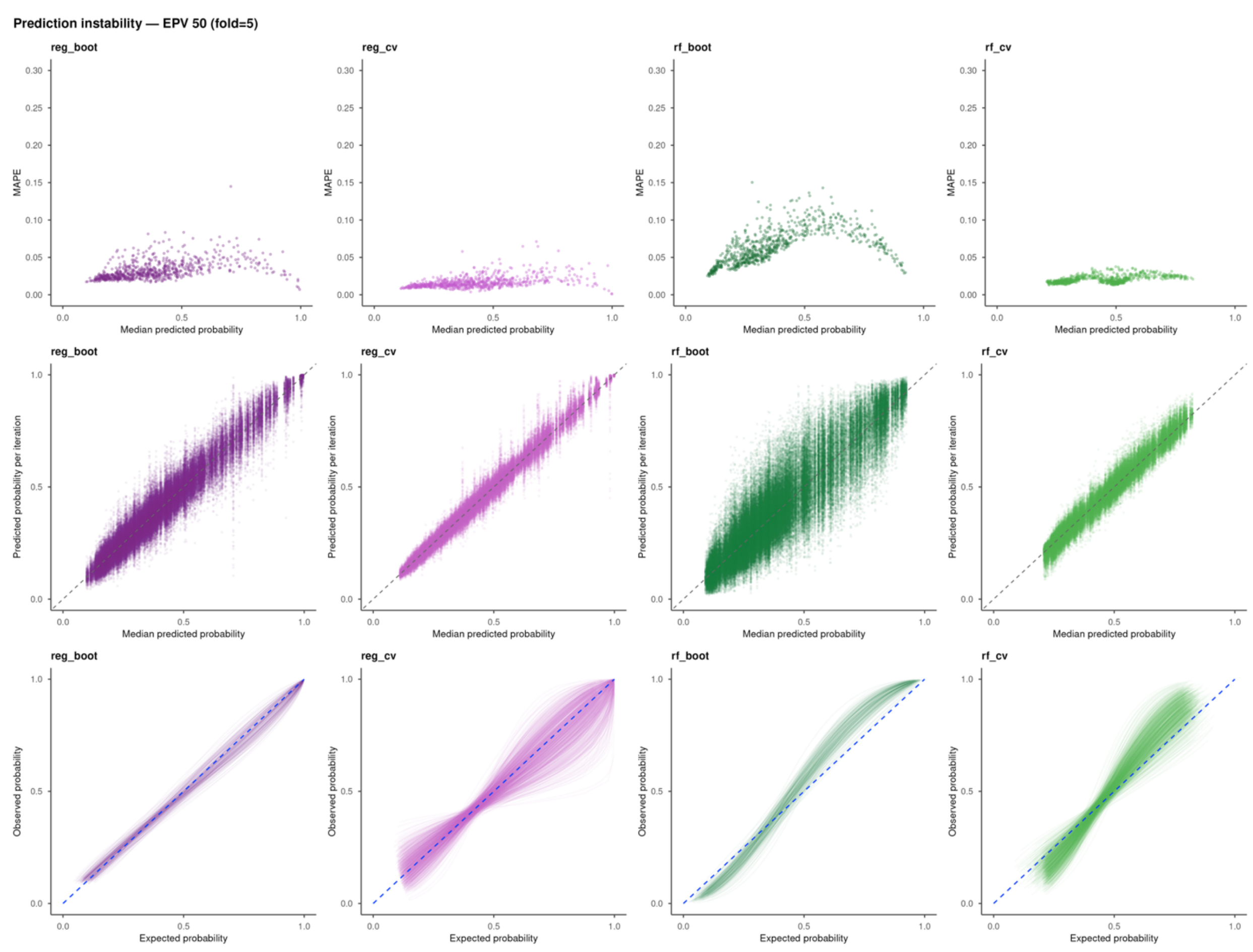

**Figure 4. Prediction instability plots at EPV 50 with 5-fold cross-validation.** Upper panels display mean absolute prediction error (MAPE) for each observation plotted against its median predicted probability across all validation iterations. Middle panels show the predicted probability from every validation iteration plotted against the median predicted probability across all iterations; the dashed diagonal line indicates perfect agreement between iteration-specific and median predictions. Lower panels illustrate calibration instability, where each line represents a logistic calibration curve derived from each individual validation iteration plotted simultaneously; the blue dashed line indicates perfect calibration. Purple and green colors represent logistic regression and random forest models, respectively. **Abbreviations**: EPV, events-per-variable;

MAPE, mean absolute prediction error; reg_boot, logistic regression with an internal validation using the bootstrap approach; reg_cv, logistic regression with an internal validation using the 5-fold cross-validation approach; rf_boot, random forest with an internal validation using the bootstrap approach; rf_cv, random forest with an internal validation using the 5-fold cross-validation approach

**Discussion**

This study proposed a repeated k-fold cross-validation (CV) approach for evaluating individual-level prediction stability alongside the discrimination and calibration performance of clinical prediction models. To examine whether this approach could provide a practical means of assessing both performance and stability within a single resampling framework, we compared it with the established bootstrap approach. We evaluated logistic regression (LR) and random forest (RF) models developed under varying events per variable (EPV) conditions: 10, 30, 50, and 1,004, the latter representing the full dataset. The two approaches yielded similar estimates of predictive performance, but characterised prediction instability in different ways.

CV is generally recognised as a robust method for internal validation of clinical prediction models. From our findings, CV and bootstrap produced similar estimates of discrimination and calibration for LR modelling, particularly at higher EPV. In contrast, for RF, discrimination was comparable between the two approaches, whereas the CV-derived calibration slope remained closer to the reference value than the bootstrap-corrected slope, indicating greater variance with the bootstrap approach. These results are consistent with previous studies. In a large clinical dataset with a rare outcome, Coley et al. (18) found that full-sample CV performance matched performance in unseen data, whereas bootstrap optimism-corrected performance was higher. Zhang et al. (19) reported that, for machine learning models, k-fold CV produced stable and approximately unbiased performance estimates, whereas bootstrap optimism-corrected estimates tended to be higher. Bradshaw et al. (20) similarly recommended CV for obtaining more stable and less optimistic performance estimates in machine learning applications in medical imaging. Taken together, these findings suggest that CV may be preferable, at least for random forest and other tree-based algorithms, and for models requiring hyperparameter tuning, whereas the bootstrap remains appropriate for regression models. However, good predictive performance does not necessarily indicate good prediction stability (24).

In our view, the choice between CV and bootstrap has greater implications for assessing prediction instability than for estimating average predictive performance. The bootstrap may be well suited to regression models, where a single prespecified model is fitted and no hyperparameter tuning is required. By contrast, CV may be better suited to machine learning algorithms(19, 20), such as RF, and to models requiring hyperparameter tuning, because it can mimic the full modelling pipeline: model development, tuning, refitting, and evaluation are repeated within each resample. In our study, each dataset was split into an 80% development sample and a 20% reference dataset, allowing performance and stability to be assessed against predictions made on previously unseen data. The 5-fold CV mirrors this 80:20 structure within the development sample, as each iteration trains the model on 80% of the data and evaluates it on the remaining 20%. The bootstrap does not reproduce this structure, because model fitting and evaluation are based on overlapping samples rather than distinct training and assessment partitions. CV may therefore better approximate the way prediction models, particularly tuned or machine learning models, are developed and evaluated in practice.

Prediction instability, as quantified by MAPE, was consistently higher with the bootstrap approach than with CV across all EPV scenarios. At an EPV of 30, for example, the median MAPE was 0.042 for bootstrap and 0.020 for CV in the LR models, and 0.077 and 0.027, respectively, in the RF models. This difference is likely explained partly by the number of distinct individuals contributing to each fitted model. A bootstrap resample of the full sample size, drawn with replacement, contains on average about 63.2% unique observations, with the remaining observations appearing as repeated draws, as noted by Efron and Tibshirani (25) and Kohavi (26). Each bootstrap model is therefore fitted on fewer distinct individuals, which can make predictions vary more across resamples. By contrast, five-fold CV fits each model on 80% distinct observations and distributes the data more evenly across folds, in line with Kohavi (26) and Molinaro et al. (27). The higher bootstrap-derived MAPE may therefore partly reflect the mechanics of fitting models on fewer distinct observations, rather than greater underlying prediction instability alone.

Beyond this intrinsic bootstrap property, bootstrap- and CV-derived MAPE are different estimators that target different estimands. Bootstrap-derived MAPE quantifies how much an individual prediction varies across resampled models relative to the prediction from the single model fitted to the original development sample. CV-derived MAPE quantifies how much an individual prediction varies across repeats relative to its own median prediction on held-out data. The two measures therefore assess instability against different reference points and in differently constituted samples, so part of the numerical gap reflects differences in the questions they answer rather than greater underlying instability. The number of distinct observations available to each model also helps explain why RF was more unstable at low EPV. Random forest depends strongly on the sampled observations and requires many more events per variable than regression to stabilise, as shown by van der Ploeg et al. (28).

The CV-based calibration instability results, however, did not follow the same pattern as the MAPE results. At low EPV, the calibration curves appeared noisier under CV than under the bootstrap, despite more stable individual predictions under CV, as shown by the MAPE and prediction instability plots. This reversal reflects the amount of data available to estimate each calibration curve. Under five-fold CV, each curve is estimated in a single held-out fold, representing about one fifth of the development sample, and is therefore based on relatively few observations and events. Under the bootstrap, each curve is estimated in the full development sample. In estimand terms, the CV curves describe calibration in unseen held-out data, whereas the bootstrap curves describe calibration in the development sample. The noisier CV curves therefore reflect imprecision from smaller held-out samples, rather than greater underlying miscalibration. This is consistent with the recognised unreliability of calibration curves in small samples, as discussed by Austin and Steyerberg (29) and Van Calster et al. (30). By contrast, MAPE and prediction instability plots pool information across individuals and repeats, making them less sensitive to the size of any single held-out fold. Thus, while CV-based assessment appears useful for individual-level prediction instability, CV-based calibration instability is more sample-size dependent and may require larger development datasets than bootstrap-based assessment. This also supported our choice of 5-fold CV, as using more folds would further reduce the size of each held-out fold and add noise to the calibration curves.

A key strength of this study is that it offers an empirical evaluation of repeated k-fold CV as a feasible approach for assessing prediction instability. This approach integrates the assessment of predictive performance and prediction instability within a single resampling framework that can reflect the full model development process, including hyperparameter tuning. This may be particularly useful when comparing multiple algorithms, as each algorithm can be evaluated using the same resampling structure. The approach is also relevant to machine learning algorithms and other models requiring tuning, for which CV may align more closely with the intended modelling workflow than the standard bootstrap approach.

Several limitations should be weighed against these strengths. Although empirically testing and comparing the two approaches in a real-world clinical dataset may better reflect applied prediction modelling practice than a purely simulation-based study, it limited our ability to examine several methodological issues in depth. The data came from a single clinical cohort with one binary outcome and six continuous predictors assumed to have linear effects, so the findings may not extend to time-to-event outcomes, categorical predictors, nonlinear effects, or more complex predictor structures. Unlike simulation studies, real-world data do not provide a known underlying truth, which prevented us from directly separating estimation error, sampling variability, and underlying prediction instability. We evaluated only one machine learning algorithm, and whether the behaviour observed for RF extends to gradient boosting, support vector machines, neural networks, or other algorithms remains to be tested, although the approach is applicable in principle to any algorithm. The mechanisms we propose, particularly for the calibration findings, should therefore be interpreted as plausible explanations rather than effects tested directly with fold-level diagnostics. For these reasons, the comparisons should be regarded as exploratory, and the findings should be confirmed across other algorithms, outcomes, sample sizes, and clinical settings before the approach is adopted in routine practice.

## Conclusions

We propose repeated cross-validation as a practical alternative to the bootstrap for evaluating prediction instability in clinical prediction models, while also obtaining internally validated estimates of predictive performance from the same analysis. In this initial evaluation, CV-derived predictive performance was similar to bootstrap-derived performance for LR, and the CV-derived calibration slope remained closer to the reference value than the bootstrap-corrected slope for RF. CV-derived MAPE was consistently lower, reflecting differences in the quantities estimated by the two approaches rather than necessarily indicating lower underlying instability. Based on this feasibility-level evidence from a single dataset, repeated cross-validation appears to be a coherent option for jointly assessing predictive performance and prediction instability in clinical prediction models. Further validation across algorithms, outcomes, sample sizes, and clinical settings is needed before it can be recommended for routine use.

## List of abbreviations

AUC, area under the receiver operating characteristic curve; BT, body temperature; CITL, calibration-in-the-large; CPM, clinical prediction model; CV, cross-validation; ED, emergency department; EPV, events per variable; HPC, high-performance computing; LR, logistic regression; MAP, mean arterial pressure; MAPE, mean absolute prediction error; OOB, out-of-bag; PR, pulse rate; RF, random forest; RR, respiratory rate; SpO2, peripheral oxygen saturation; TRIPOD, Transparent Reporting of a multivariable prediction model for Individual Prognosis Or Diagnosis.

## Declarations

### Ethics approval and consent to participate

This study was exempted from ethics review by the ethical committee of the Faculty of Medicine at Chiang Mai University (certificate of exemption no. 0267/2025) as it involved secondary analysis of de-identified data from a previously approved clinical prediction model development study.

### Consent for publication

Not applicable.

### Availability of data and materials

The dataset analysed during the current study is available from the corresponding author on reasonable request. The simulation code and analysis scripts are available at https://github.com/Nopkhongthon/predinstab.

### Competing interests

The authors declare that they have no competing interests.

**Funding**

This research received no specific funding from any agency in the public, commercial, or not-for-profit sectors.

**Authors' contributions**

**Nop Khongthon**: Conceptualisation, Methodology, Software, Formal analysis, Investigation, Data curation, Validation, Visualisation, Writing - Original Draft, Writing - Review & Editing. **Phichayut Phinyo**: Conceptualisation, Methodology, Software, Formal analysis, Investigation, Data curation, Validation, Visualisation, Writing - Original Draft, Writing - Review & Editing, Supervision, Project administration. **Pakpoom Wongyikul**: Conceptualisation, Methodology, Validation. **Noraworn Jirattikanwong**: Conceptualisation, Methodology, Validation. **Phanu Prasankittirach**: Conceptualisation, Methodology, Validation. **Natthanaphop Isaradech**: Conceptualisation, Methodology, Validation. **Wachiranun Sirikul**: Conceptualisation, Methodology, Validation, Supervision, Project administration. **Noppadon Seesuwan**: Dataset. **Suppachai Lawanaskol**: Dataset. All authors read and approved the final manuscript.

**Acknowledgements**

This study was partially supported by the Faculty of Medicine, Chiang Mai University. The authors also acknowledge the use of the high-performance computing resource, Raptor, at the Faculty of Medicine, Chiang Mai University.

**Additional files**

Additional file 1: Supplementary Figures S1–S11. Prediction instability plots for logistic regression and random forest models, presented across events-per-variable conditions (S1–S4), sample size scenarios by modelling and validation approach (S5–S8), and modelling approaches by events-per-variable condition (S9–S11). File format: DOCX.

## Additional file 1: Supplementary Figures S1–S11.

**Supplementary Figure S1.** Prediction instability at EPV 10 with 5-fold cross-validation

**Supplementary Figure S2.** Prediction instability at EPV 30 with 5-fold cross-validation

**Supplementary Figure S3.** Prediction instability at EPV 50 with 5-fold cross-validation

**Supplementary Figure S4.** Prediction instability at EPV full with 5-fold cross-validation

**Supplementary Figure S5.** Prediction instability across sample size scenarios for logistic regression with bootstrap method

**Supplementary Figure S6.** Prediction instability across sample size scenarios for logistic regression with 5-fold cross-validation

**Supplementary Figure S7.** Prediction instability across sample size scenarios for random forest with bootstrap method

**Supplementary Figure S8.** Prediction instability across sample size scenarios for random forest with cross-validation

**Supplementary Figure S9.** Prediction instability across modelling approaches at EPV 10 with 5-fold cross-validation

**Supplementary Figure S10.** Prediction instability across modelling approaches at EPV 30 with 5-fold cross-validation

**Supplementary Figure S11.** Prediction instability across modelling approaches at EPV full with 5-fold cross-validation

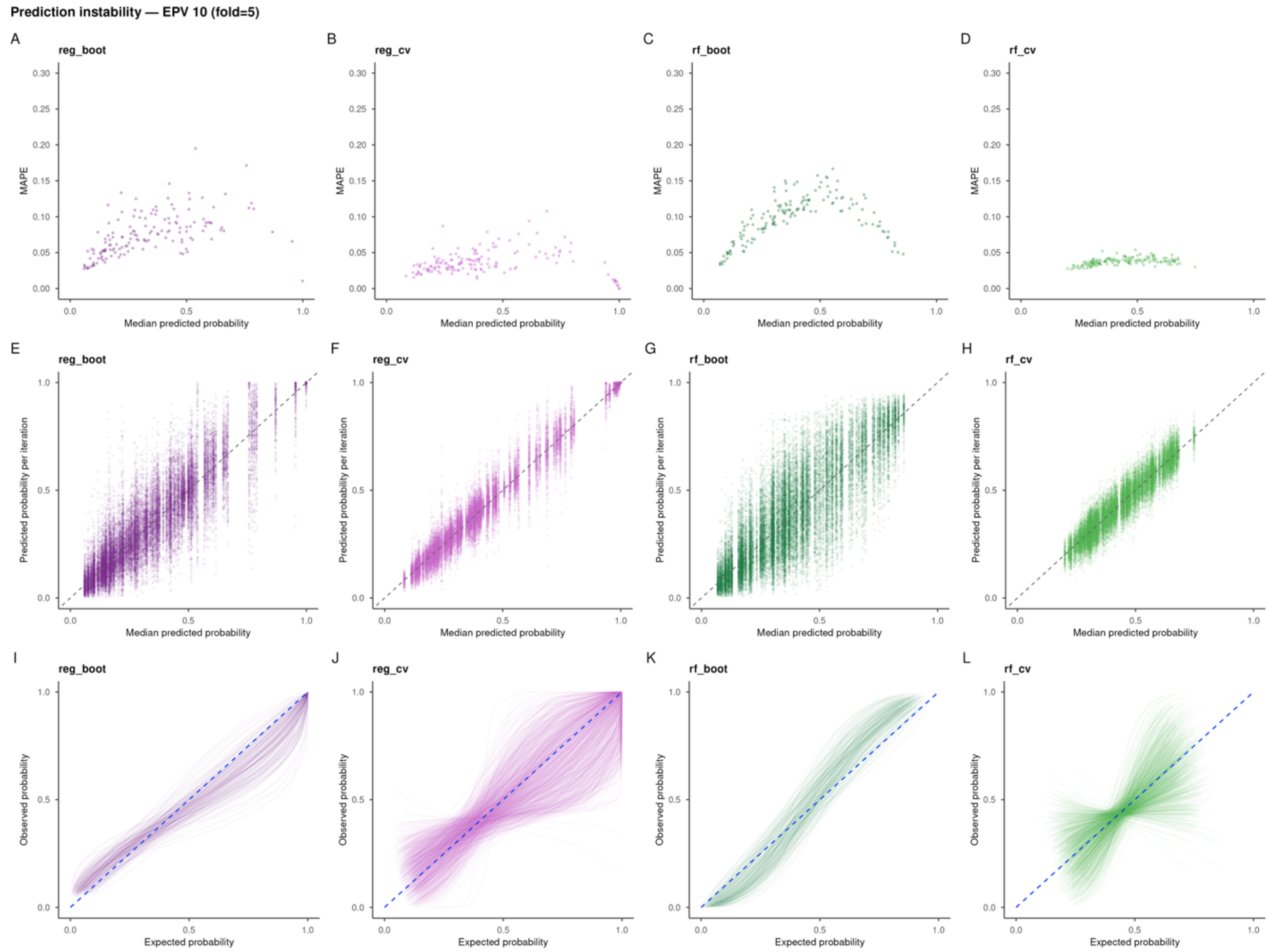

**Supplementary Figure S1. Prediction instability at EPV 10 with 5-fold cross-validation.** Each row displays a different instability metric. The top row shows MAPE, with each point representing an individual observation. The middle row shows prediction instability plots of bootstrap- or cross-validation-derived predicted probabilities against the median predicted probability for each individual. The bottom row shows calibration instability curves of observed against expected predicted probabilities across iterations. Columns correspond to the four modelling approaches. Purple and green points represent logistic regression and random forest models, respectively. **Abbreviations**: EPV, events-per-variable; MAPE, mean absolute prediction error; reg_boot, logistic regression with bootstrap method; reg_cv, logistic regression with 5-fold cross-validation; rf_boot, random forest with bootstrap method; rf_cv, random forest with cross-validation.

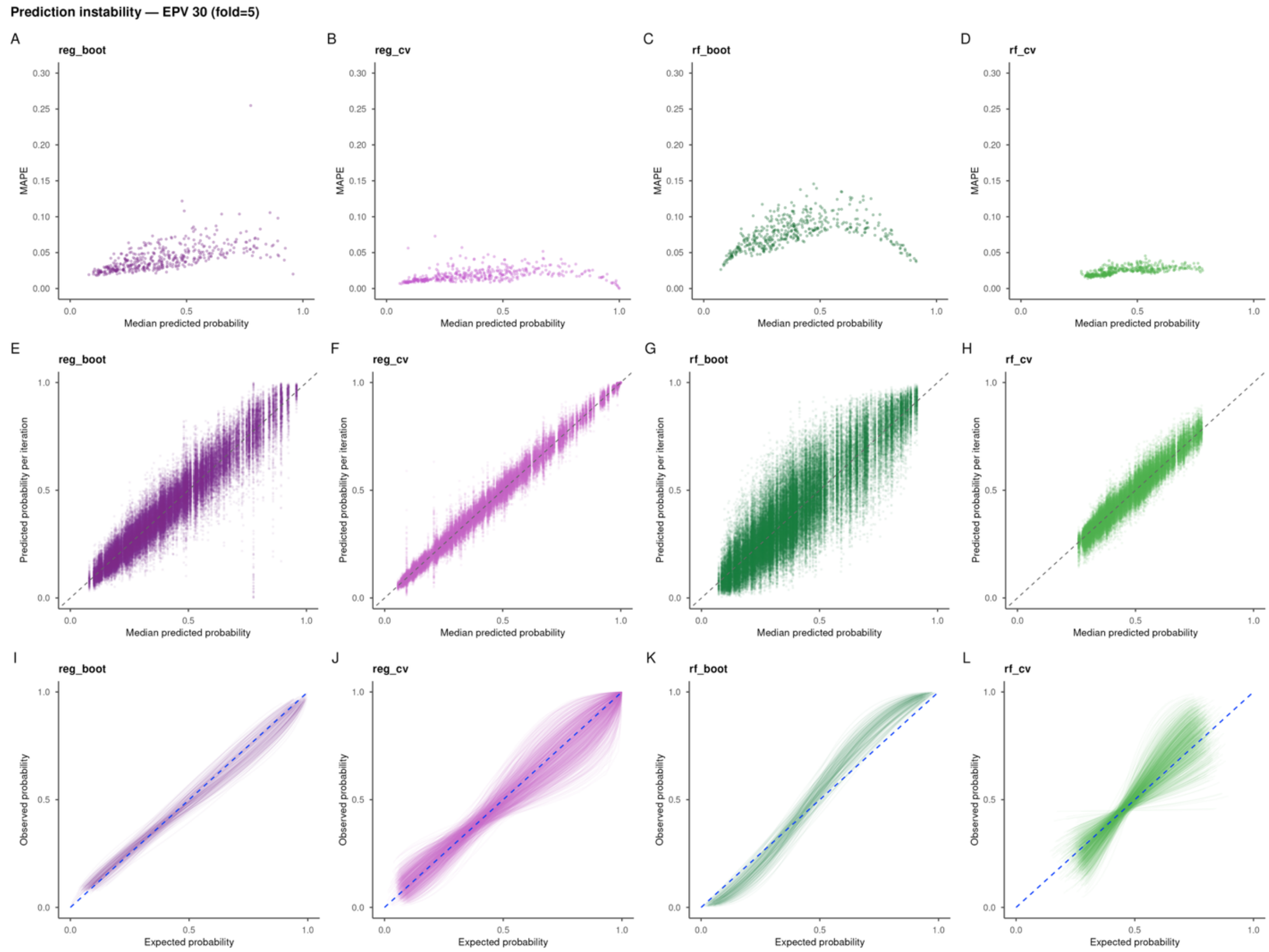

**Supplementary Figure S2. Prediction instability at EPV 30 with 5-fold cross-validation.** Each row displays a different instability metric. The top row shows MAPE, with each point representing an individual observation. The middle row shows prediction instability plots of bootstrap- or cross-validation-derived predicted probabilities against the median predicted probability for each individual. The bottom row shows calibration instability curves of observed against expected predicted probabilities across iterations. Columns correspond to the four modelling approaches. Purple and green points represent logistic regression and random forest models, respectively. **Abbreviations**: EPV, events-per-variable; MAPE, mean absolute prediction error; reg_boot, logistic regression with bootstrap method; reg_cv, logistic regression with 5-fold cross-validation; rf_boot, random forest with bootstrap method; rf_cv, random forest with cross-validation.

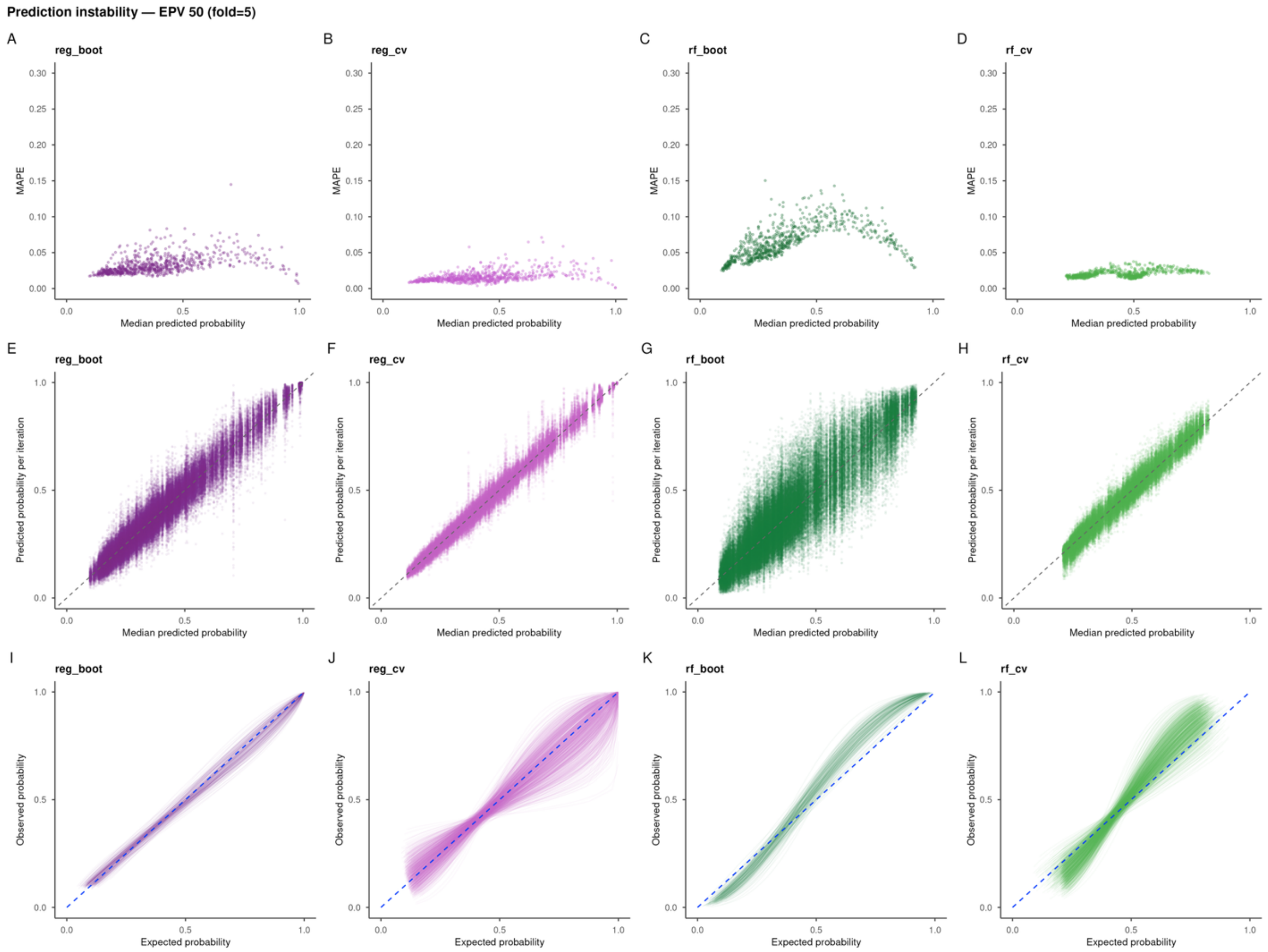

**Supplementary Figure S3. Prediction instability at EPV 50 with 5-fold cross-validation.** Each row displays a different instability metric. The top row shows MAPE, with each point representing an individual observation. The middle row shows prediction instability plots of bootstrap- or cross-validation-derived predicted probabilities against the median predicted probability for each individual. The bottom row shows calibration instability curves of observed against expected predicted probabilities across iterations. Columns correspond to the four modelling approaches. Purple and green points represent logistic regression and random forest models, respectively. **Abbreviations**: EPV, events-per-variable; MAPE, mean absolute prediction error; reg_boot, logistic regression with bootstrap method; reg_cv, logistic regression with 5-fold cross-validation; rf_boot, random forest with bootstrap method; rf_cv, random forest with cross-validation.

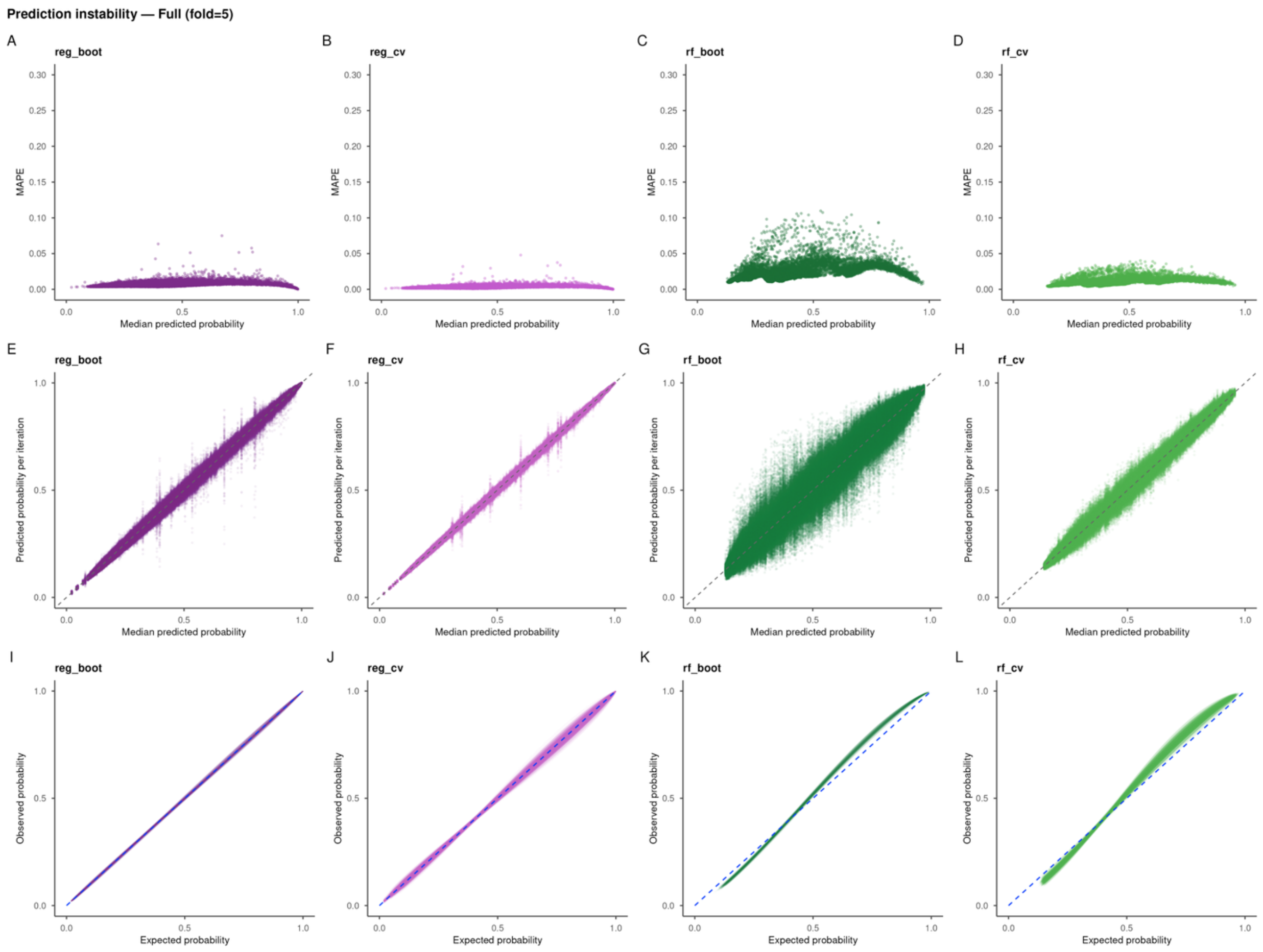

**Supplementary Figure S4. Prediction instability at EPV full with 5-fold cross-validation.** Each row displays a different instability metric. The top row shows MAPE, with each point representing an individual observation. The middle row shows prediction instability plots of bootstrap- or cross-validation-derived predicted probabilities against the median predicted probability for each individual. The bottom row shows calibration instability curves of observed against expected predicted probabilities across iterations. Columns correspond to the four modelling approaches. Purple and green points represent logistic regression and random forest models, respectively. **Abbreviations**: EPV, events-per-variable; MAPE, mean absolute prediction error; reg_boot, logistic regression with bootstrap method; reg_cv, logistic regression with 5-fold cross-validation; rf_boot, random forest with bootstrap method; rf_cv, random forest with cross-validation.

## Supplementary

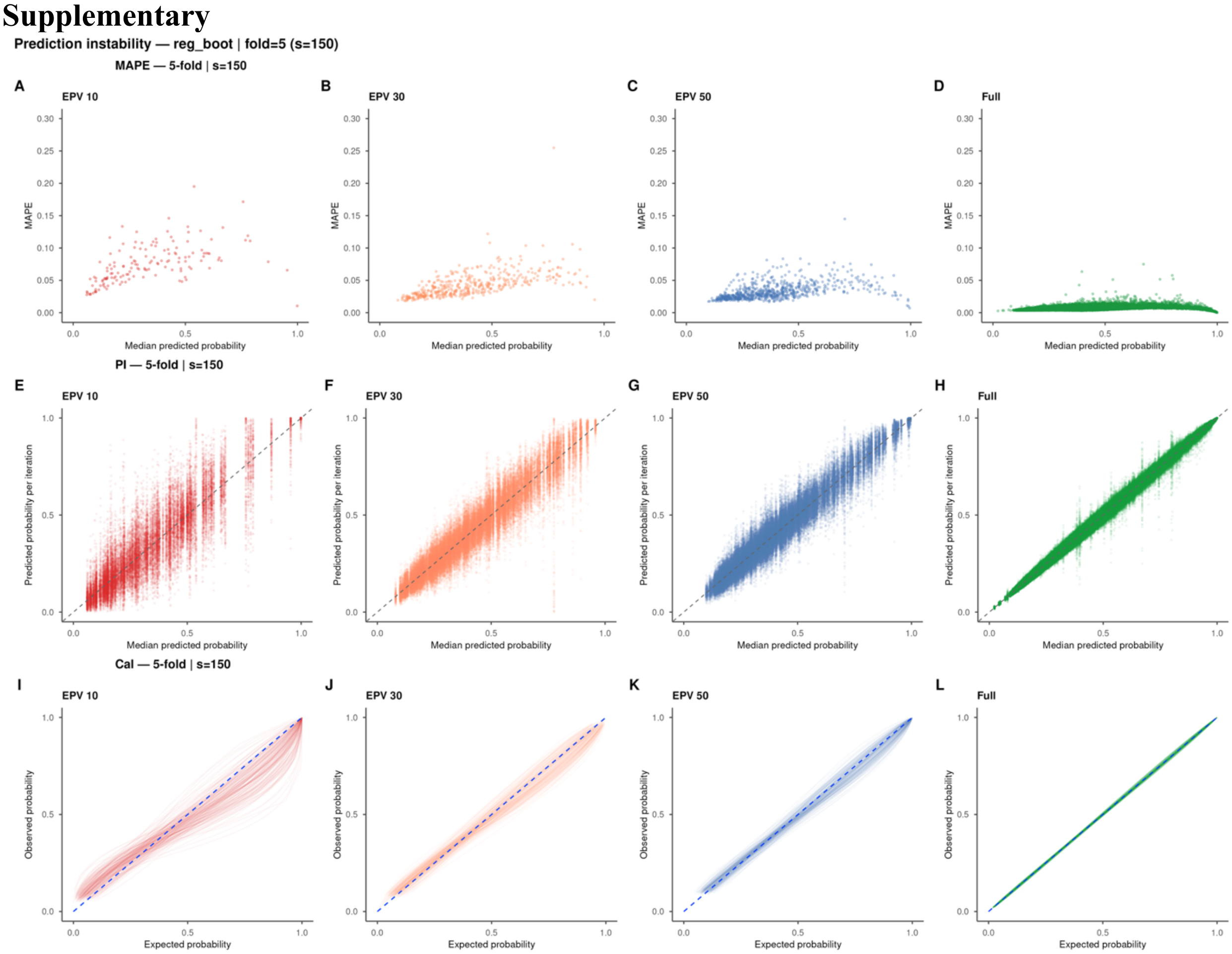

**Figure S5. Prediction instability across sample size scenarios for logistic regression with bootstrap method.** Each row displays a different instability metric across four sample size scenarios (EPV 10, EPV 30, EPV 50, and full dataset). The top row shows MAPE, with each point representing an individual observation. The middle row shows prediction instability plots of bootstrap-derived predicted probabilities against the <u>median predicted probability</u> for each individual. The bottom row shows calibration instability curves of iterated predicted probabilities against the original predicted probability. Instability decreased markedly as sample size increased, with the full dataset showing near-perfect agreement between bootstrap and reference predicted probabilities. **Abbreviations:** Cal, calibration instability curve; EPV, events-per-variable; MAPE, mean absolute prediction error; PI, prediction instability plot; reg_boot, logistic regression with bootstrap method.

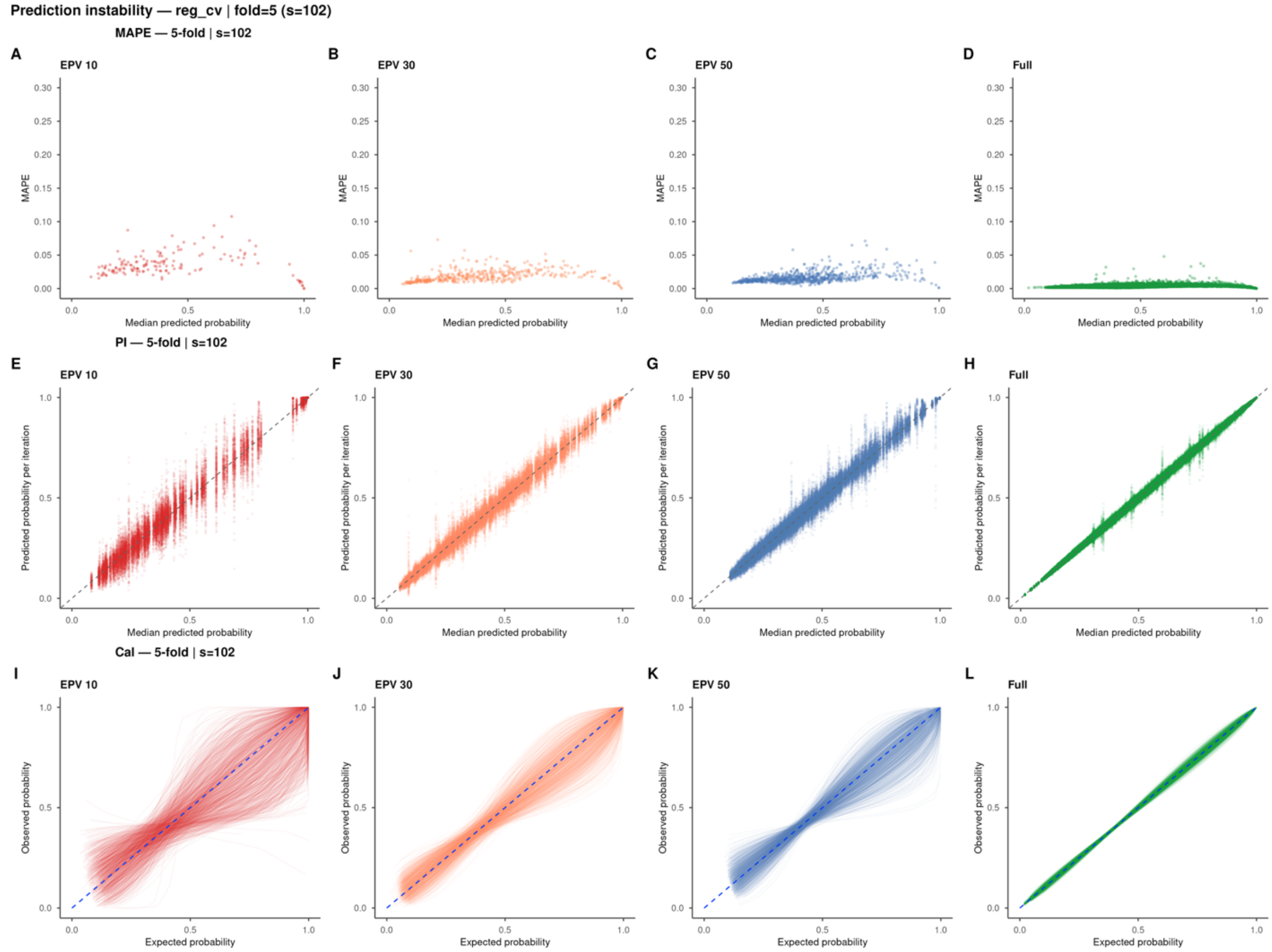

**Supplementary Figure S6. Prediction instability across sample size scenarios for logistic regression with 5-fold cross-validation.** Each row displays a different instability metric across four sample size scenarios (EPV 10, EPV 30, EPV 50, and full dataset). The top row shows MAPE, with each point representing an individual observation. The middle row shows prediction instability plots of cross-validation-derived predicted probabilities against the median predicted probability for each individual. The bottom row shows calibration instability curves of iterated predicted probabilities against the original predicted probability. Instability decreased as sample size increased; however, calibration instability curves at smaller sample sizes showed wider dispersion compared with the bootstrap approach, particularly at EPV 10. **Abbreviations**: Cal, calibration instability curve; EPV, events-per-variable; MAPE, mean absolute prediction error; PI, prediction instability plot; reg_cv, logistic regression with 5-fold cross-validation.

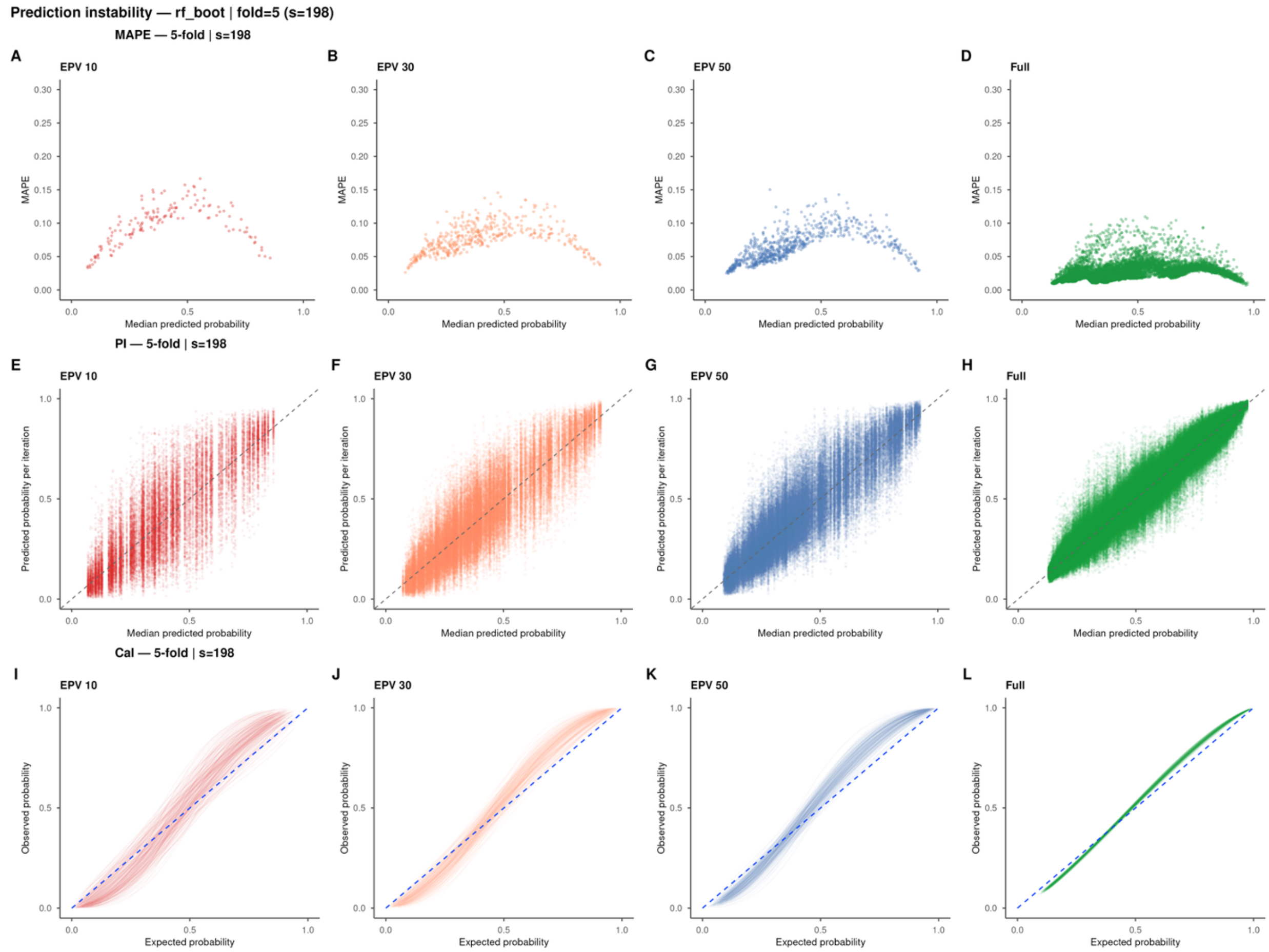

**Supplementary Figure S7. Prediction instability across sample size scenarios for random forest with bootstrap method.** Each row displays a different instability metric across four sample size scenarios (EPV 10, EPV 30, EPV 50, and full dataset). The top row shows MAPE, with each point representing an individual observation. The middle row shows prediction instability plots of bootstrap-derived predicted probabilities against the median predicted probability for each individual. The bottom row shows calibration instability curves of iterated predicted probabilities against the original predicted probability. Compared with logistic regression, random forest exhibited greater spread in prediction instability plots across all sample sizes, with wider dispersion persisting even at larger sample sizes, reflecting higher overall instability. **Abbreviations**: Cal, calibration instability curve; EPV, events-per-variable; MAPE, mean absolute prediction error; PI, prediction instability plot; rf_boot, random forest with bootstrap method.

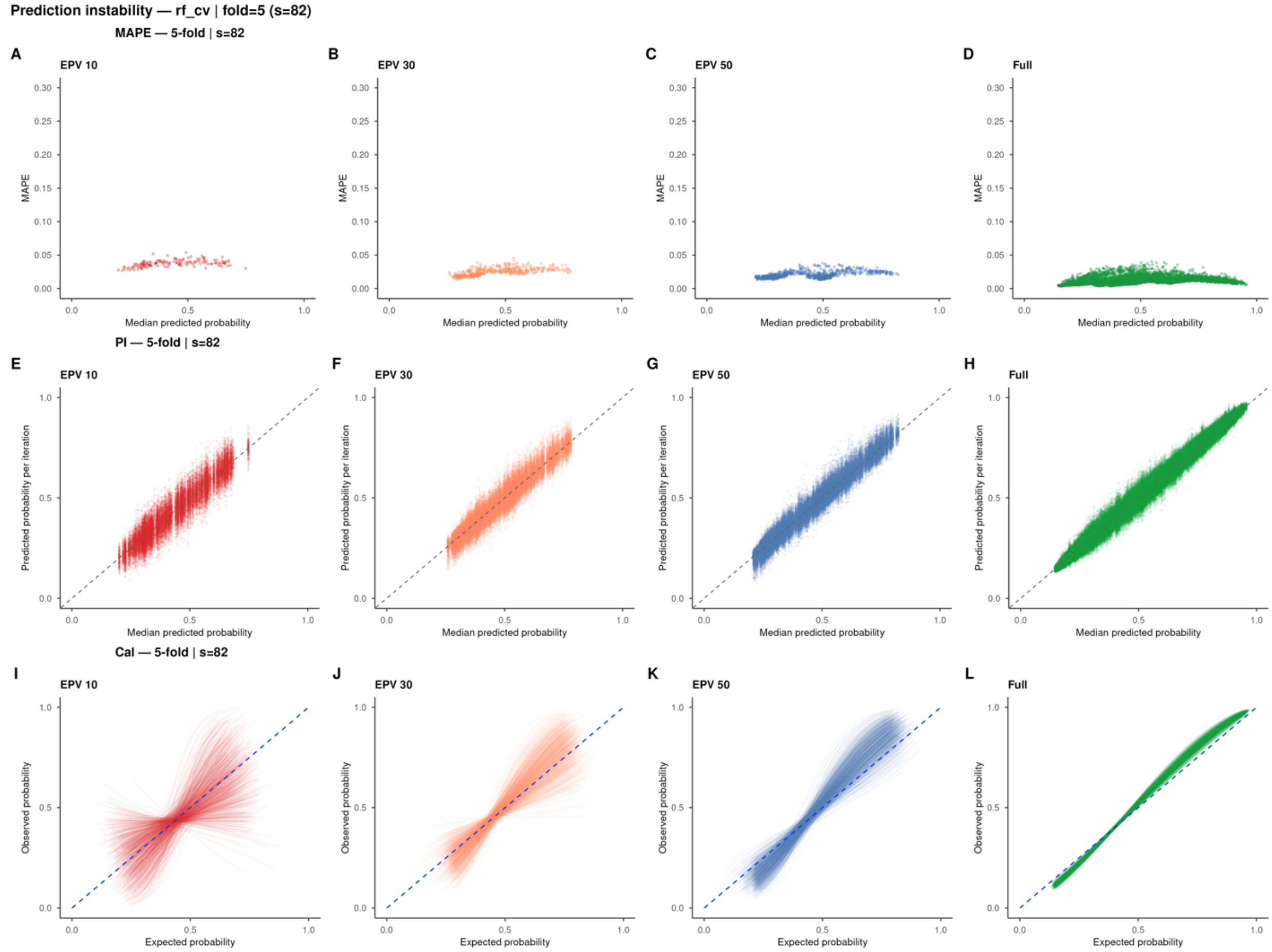

**Supplementary Figure S8. Prediction instability across sample size scenarios for random forest with cross-validation.** Each row displays a different instability metric across four sample size scenarios (EPV 10, EPV 30, EPV 50, and full dataset). The top row shows MAPE, with each point representing an individual observation. The middle row shows prediction instability plots of cross-validation-derived predicted probabilities against the median predicted probability for each individual. The bottom row shows calibration instability curves of iterated predicted probabilities against the original predicted probability. Random forest with cross-validation demonstrated notably lower MAPE and tighter prediction instability plots compared with the bootstrap approach, particularly at smaller sample sizes; however, calibration instability curves at EPV 10 showed a characteristic crossing pattern, suggesting systematic miscalibration under repeated cross-validation at low sample sizes. **Abbreviations:** Cal, calibration instability curve; EPV, events-per-variable; MAPE, mean absolute prediction error; PI, prediction instability plot; rf_cv, random forest with cross-validation.

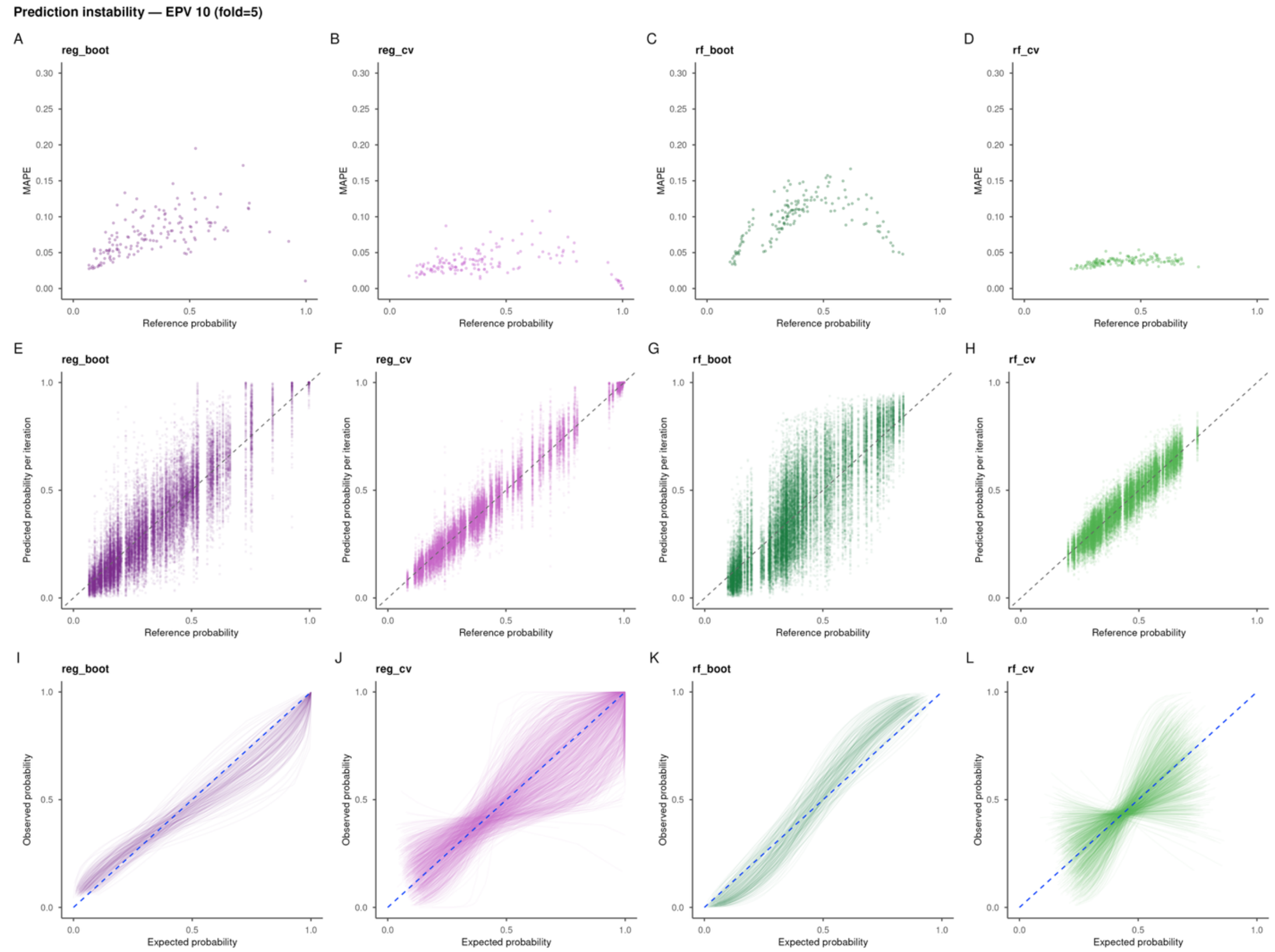

**Supplementary Figure S9. Prediction instability across modelling approaches at EPV 10 with 5-fold cross-validation.** Each row displays a different instability metric across four modelling approaches. The top row shows MAPE per individual plotted against the median predicted probability. The middle row shows prediction instability plots of iteration-specific predicted probabilities against the reference predicted probability for each individual, where the reference predicted probability is defined as the original predicted probability for bootstrap-based approaches and the median predicted probability across repeated cross-validation for cross-validation-based approaches. The bottom row shows calibration instability curves of observed against expected predicted probabilities across iterations. Columns correspond to the four modelling approaches. Purple and green colours represent logistic regression and random forest models, respectively. **Abbreviations**: EPV, events per variable; MAPE, mean absolute prediction error; reg_boot, logistic regression with bootstrap internal validation; reg_cv, logistic regression with 5-fold cross-validation; rf_boot, random forest with bootstrap method; rf_cv, random forest with cross-validation.

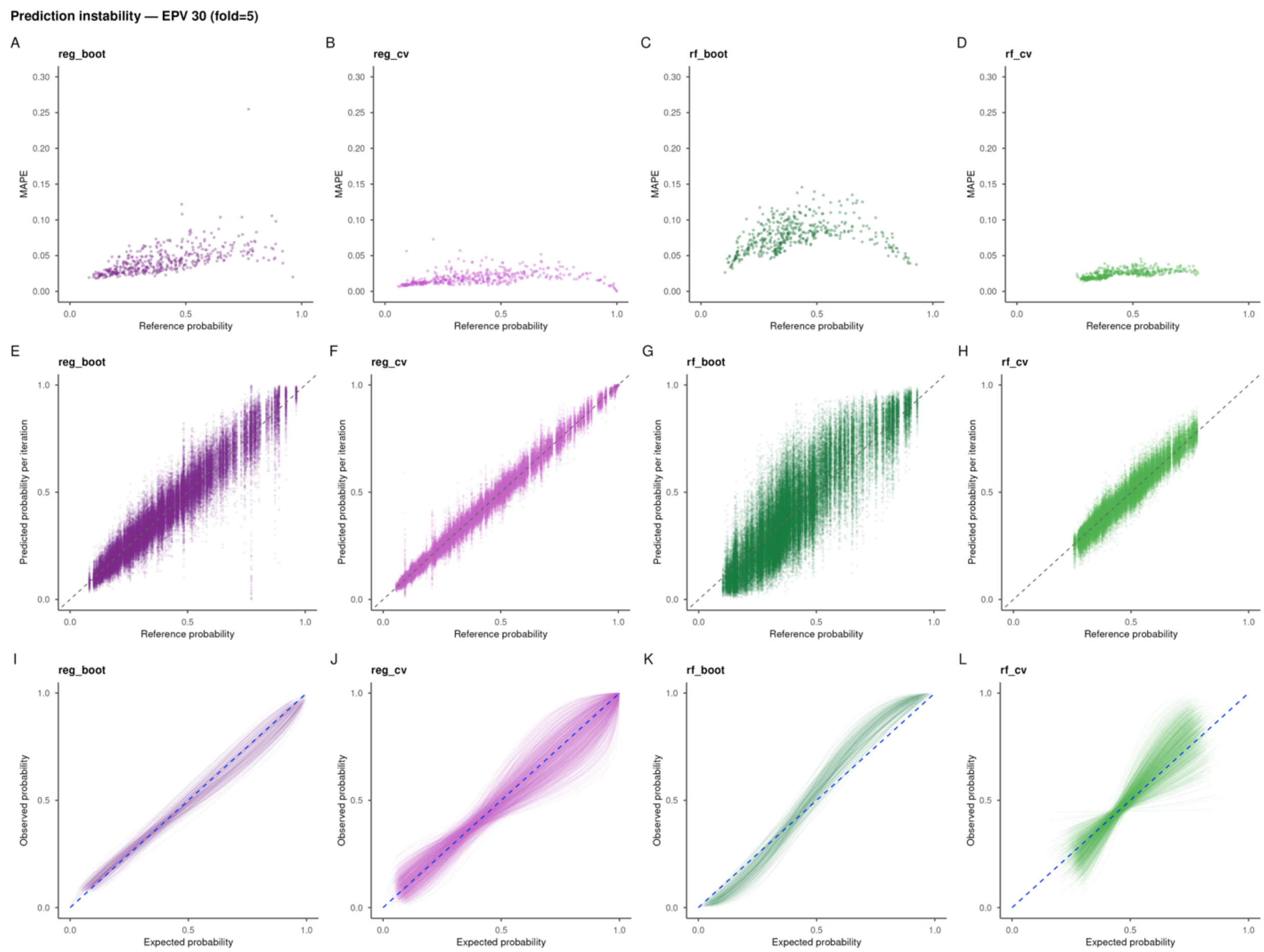

**Supplementary Figure S10. Prediction instability across modelling approaches at EPV 30 with 5-fold cross-validation.** Each row displays a different instability metric across four modelling approaches. The top row shows MAPE per individual plotted against the median predicted probability. The middle row shows prediction instability plots of iteration-specific predicted probabilities against the reference predicted probability for each individual, where the reference predicted probability is defined as the original predicted probability for bootstrap-based approaches and the median predicted probability across repeated cross-validation for cross-validation-based approaches. The bottom row shows calibration instability curves of observed against expected predicted probabilities across iterations. Columns correspond to the four modelling approaches. Purple and green colours represent logistic regression and random forest models, respectively. **Abbreviations**: EPV, events per variable; MAPE, mean absolute prediction error; reg_boot, logistic regression with bootstrap internal validation; reg_cv, logistic regression with 5-fold cross-validation; rf_boot, random forest with bootstrap method; rf_cv, random forest with cross-validation.

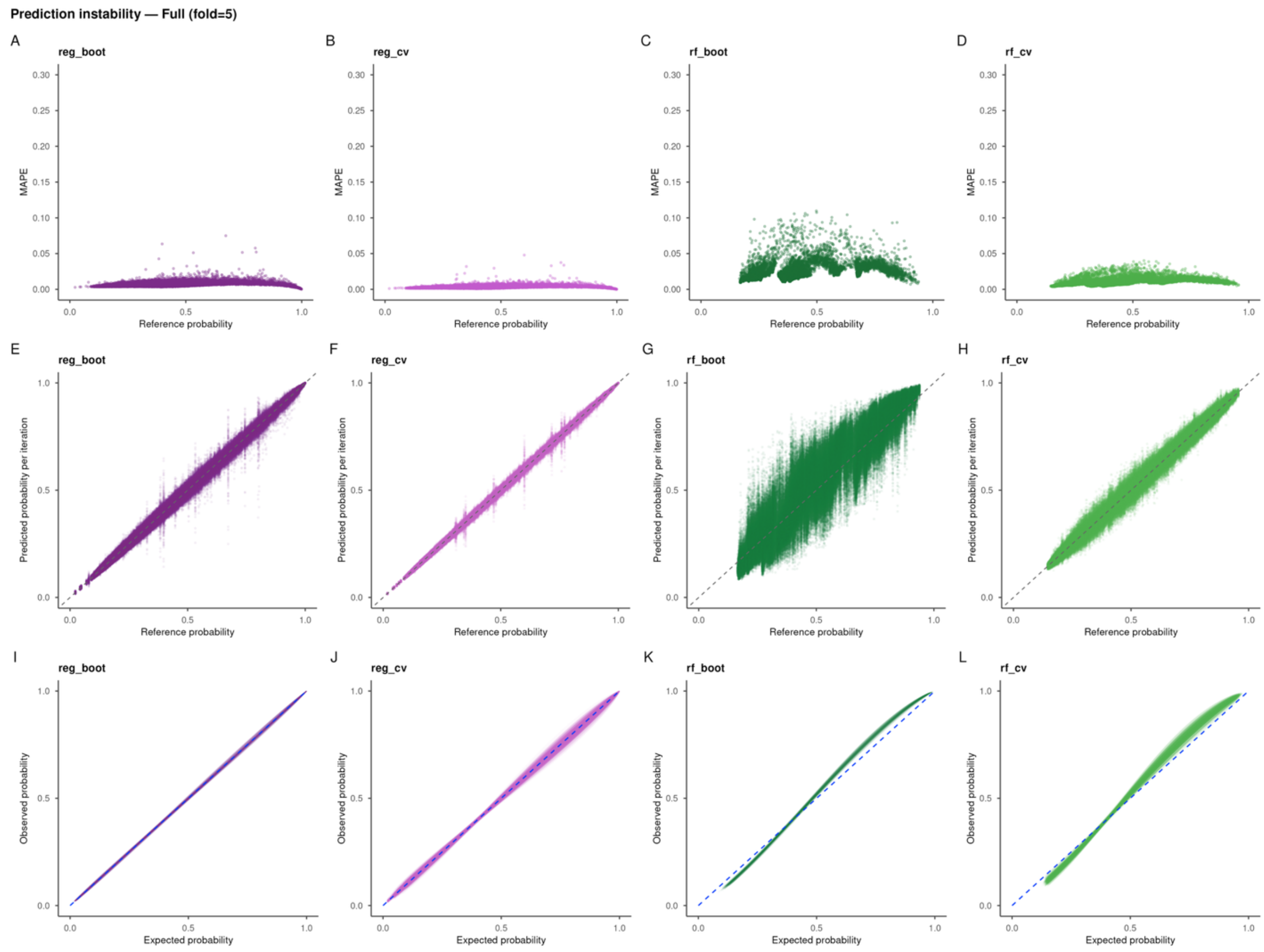

**Supplementary Figure S11. Prediction instability across modelling approaches at EPV full with 5-fold cross-validation.** Each row displays a different instability metric across four modelling approaches. The top row shows MAPE per individual plotted against the median predicted probability. The middle row shows prediction instability plots of iteration-specific predicted probabilities against the reference predicted probability for each individual, where the reference predicted probability is defined as the original predicted probability for bootstrap-based approaches and the median predicted probability across repeated cross-validation for cross-validation-based approaches. The bottom row shows calibration instability curves of observed against expected predicted probabilities across iterations. Columns correspond to the four modelling approaches. Purple and green colours represent logistic regression and random forest models, respectively. **Abbreviations**: EPV, events per variable; MAPE, mean absolute prediction error; reg_boot, logistic regression with bootstrap internal validation; reg_cv, logistic regression with 5-fold cross-validation; rf_boot, random forest with bootstrap method; rf_cv, random forest with cross-validation.